\documentclass[sigconf,9pt]{acmart}

\acmConference[ICCAD '24]{Design Automation Conference}{June 23--27, 2024}{San Francisco, CA}

\setcopyright{none}

\usepackage{multirow}
\usepackage{subfigure}
\usepackage{bbding}
\usepackage{mathtools}

\usepackage{algorithm}
\usepackage{algpseudocode}
\algtext*{EndFor}
\algtext*{EndFunction}
\algtext*{Until}
\algtext*{EndIf}
\algtext*{EndWhile}
\usepackage{amsmath}
\usepackage{array}
\usepackage[referable]{threeparttablex}
\usepackage{tablefootnote}

\usepackage{geometry}
\AtBeginDocument{%
  \providecommand\BibTeX{{%
    \normalfont B\kern-0.5em{\scshape i\kern-0.25em b}\kern-0.8em\TeX}}}

\copyrightyear{2025} 
\acmYear{2025} 
\setcopyright{acmlicensed}\acmConference[ASPDAC '25]{30th Asia and South
Pacific Design Automation Conference}{January 20--23, 2025}{Tokyo, Japan}
\acmBooktitle{30th Asia and South Pacific Design Automation Conference
(ASPDAC '25), January 20--23, 2025, Tokyo, Japan}
\acmDOI{10.1145/3658617.3697554}
\acmISBN{979-8-4007-0635-6/25/01}

\usepackage{tikz}
\usepackage{xcolor}
\usepackage{float}%exact location 
\usepackage{graphicx} 
\usepackage{caption} 
\usepackage{subcaption}

\begin{document}
%%%%%%%%%%%---SETME-----%%%%%%%%%%%%%

%\vspace{-.1in}
\begin{abstract}
%\vspace{-.03in}
Power efficiency is a critical design objective in modern processor design. A high-fidelity architecture-level power modeling method is greatly needed by CPU architects for guiding early optimizations. However, traditional architecture-level power models can not meet the accuracy requirement, largely due to the discrepancy between the power model and actual design implementation. While some machine learning (ML)-based architecture-level power modeling methods have been proposed in recent years, the data-hungry ML model training process requires sufficient similar known designs, which are unrealistic in many development scenarios.

This work proposes a new power modeling solution FirePower that targets few-shot learning scenario for new target architectures. FirePower proposes multiple new policies to utilize cross-architecture knowledge. First, it develops power models at component level, and components are defined in a power-friendly manner. Second, it supports different generalization strategies for models of different components. Third, it formulates generalizable and architecture-specific design knowledge into two separate models. FirePower also supports the evaluation of the generalization quality. In our experiments, FirePower can achieve a low error percentage of 5.8\% and a high correlation $R$ of 0.98 on average only using two configurations of target architecture. This is 8.8\% lower in error percentage and 0.03 higher in $R$ compared with directly training McPAT-Calib baseline on configurations of target architecture.
%\vspace{-.05in}
\end{abstract}
%\vspace{-.05in}
%\vspace{-.05in}



%\title{A Few-Shot Architecture-Level Power Modeling Framework with Cross-Architecture Knowledge Transfer}

%\title{Towards a Foundation of : An Architecture-Level Power Modeling Framework that Extracts Cross-Architecture Knowledge}

\title{FirePower: Towards a \underline{F}oundation with Generalizable Knowledge for Arch\underline{i}tectu\underline{re}-Level \underline{Power} Modeling}

\author{ 
%\fontsize{12}{12}\selectfont 
%\vspace{-.02in}
\large Qijun Zhang, Mengming Li, Yao Lu, Zhiyao Xie*\\ 
%\vspace{.05in}
%\vspace{-.01in}
\large Hong Kong University of Science and Technology\\
%\vspace{-.01in}
%\fontsize{10}{10}\selectfont 
\large \{qzhangcs, mengming.li, yludf\}@connect.ust.hk, eezhiyao@ust.hk
%\vspace{-.04in}
}

% with Common Knowledge Extraction 

%: A Cross-Architecture Solution with Common Knowledge Extraction}

%\title{A Few-Shot Architecture-level Power Modeling Framework with Common Knowledge Extraction}

%\title{A Few-Shot Cross-Architecture Power Modeling Framework}

% with Cross-Architecture Knowledge Transfer}

% automated (without designer expertise)
% few-shot
% Microarchitecture 

\maketitle
\pagestyle{plain}
\begingroup\renewcommand\thefootnote{*}
\footnotetext{Corresponding Author}
\endgroup

%of their iterative processor design. 
% prior to the RTL implementation

%Some architecture-level power modeling methods have been proposed and widely adopted, such as McPAT~\cite{li2009mcpat} and Wattch~\cite{brooks2000wattch}. However, 

%\vspace{-.1in}
\section{Introduction}
%\vspace{-.03in}
Power efficiency is a critical design objective for processor design. With the increasing design complexity, it takes significant time and effort in power optimization. As a result, a fast and highly accurate architecture-level power modeling method is greatly needed for early power evaluation prior to RTL implementation. Traditional analytical architecture-level power models such as McPAT~\cite{li2009mcpat} and Wattch~\cite{brooks2000wattch} are often inaccurate, largely due to the discrepancy among architecture-level simulator, power model, and real target CPU. This has been discussed in many prior works~\cite{xi2015quantifying,zhai2022mcpat}. Despite some works~\cite{tang2014mcpat, guler2020mcpat} updating internal design of analytical power models, they require significant human efforts and are primarily developed in-house to cater to proprietary designs.

\begin{figure}[!t]
\centering
%\hspace{-4mm}
\includegraphics[width=0.45\textwidth]{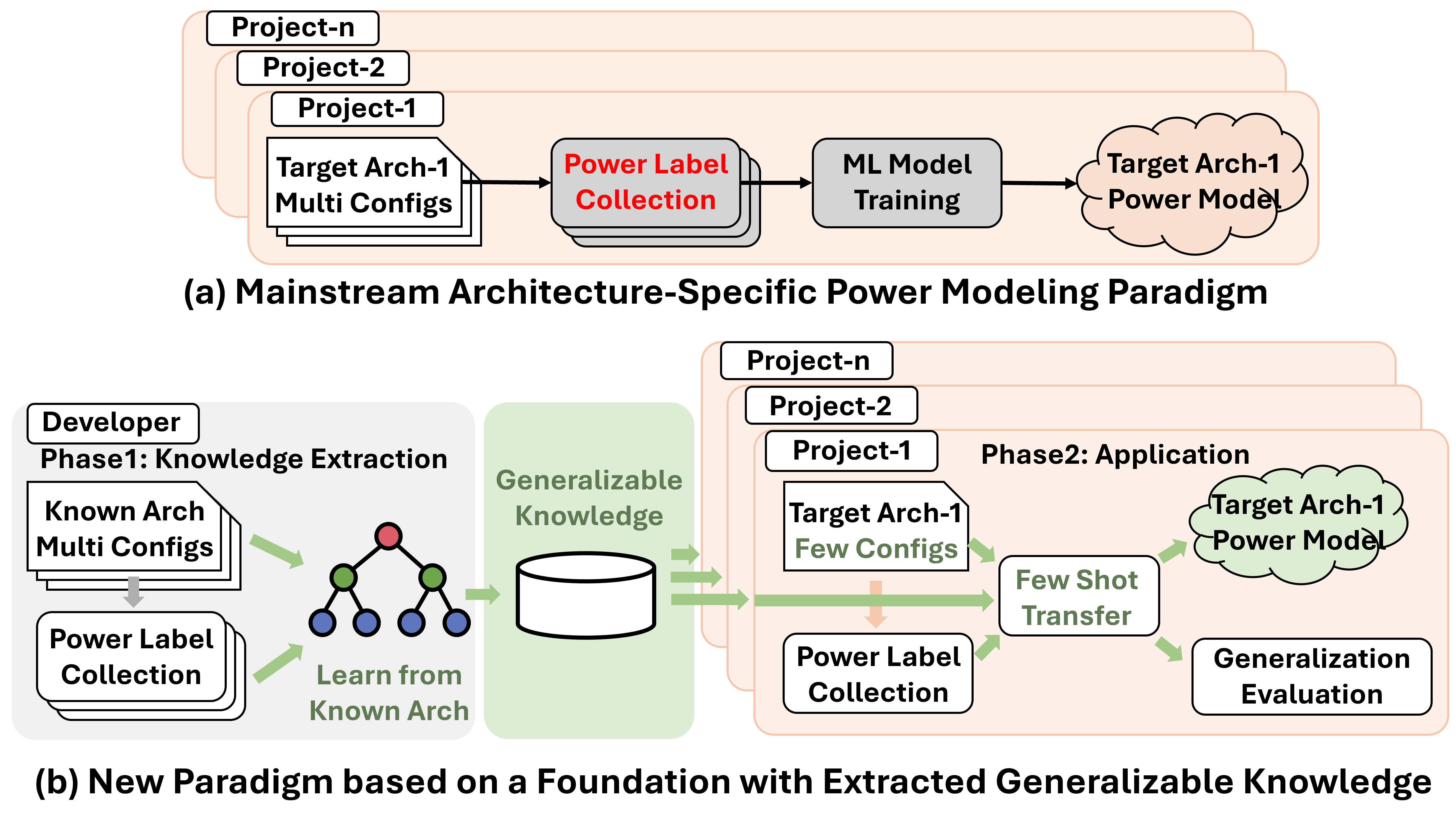}
\vspace{-5mm}
\caption{Proposed power modeling paradigm FirePower vs. existing architecture-specific paradigm. FirePower targets few-shot learning for new target architectures. It extracts general knowledge from an already known architecture, providing a ``foundation'' to support modeling new architectures.} 
\vspace{-6mm}
\label{paradigm}
\end{figure}

In recent years, ML-based architecture-level power modeling methods~\cite{lee2015powertrain,zhai2022mcpat,zhai2023microarchitecture,zhang2023panda} have been explored and demonstrated better accuracy by calibrating analytical models with ML models. However, as Fig.~\ref{paradigm}(a) shows, most ML-based power models are developed for a specific architecture, and are only applicable to new configurations under exactly the same architecture. For example, most prior works train and test power models on BOOM CPUs~\cite{zhao2020sonicboom} only. Training and testing are performed on different BOOM configurations, sharing obvious similarities. Building an ML-based power model requires sufficient ground-truth power labels of known configurations of target architecture~\cite{lee2015powertrain,zhai2022mcpat,zhai2023microarchitecture}. For a new project developing a slightly different architecture, the whole power model must be retrained from scratch based on a brand-new training dataset.

In practice, for an ongoing project targeting a specific architecture, there are not many already accomplished designs available to provide training labels. If collecting labels from scratch, the process can be highly expensive: For each design configuration, its label collection requires register-transfer level (RTL) implementation, synthesis, and simulation with workloads. RTL implementation can be especially tedious. A more practical scenario is, only a few design configurations under the target architecture are available to provide training labels. This is a typical few-shot learning scenario. The existing architecture-specific power modeling paradigm suffers from limited accuracy when only a few labels are available.

Motivated by the limitation of architecture-specific methods, we propose to extract general knowledge that can be applied across architectures. This is possible since different architectures still have similarities. Take out-of-order CPUs as an example, despite different design decisions, general design principles are similar, leading to partially similar power characteristics. However, separating general and specific knowledge at the architecture level is challenging since there is no clear standard of general or architecture-specific part of power. To the best of our knowledge, no prior ML power models have explicitly explored this topic at the architecture level.

In this work, we propose a two-phase power modeling paradigm named FirePower, as shown in Fig.~\ref{paradigm}(b). (1) The first phase is named knowledge extraction. This first phase can be time-consuming, but it is performed only once and by us, FirePower developers. Developers will first collect data on one known architecture (e.g., BOOM CPU series). Sufficient power labels will be collected for multiple configurations with good coverage for the design space. The framework will extract generalizable knowledge from this architecture. This general knowledge will provide a \emph{foundation} for the few-shot power modeling tasks on other architectures. (2) The second phase will utilize the foundation with general knowledge to build the power model for each new target architecture. With only a few available configurations of the target architecture (i.e., few-shot), the power model significantly outperforms existing architecture-specific models, which have to be trained from scratch.

The contributions of this work can be summarized below. 
\begin{itemize}
    \item We analyze the limitation of the architecture-specific power modeling paradigm when applied to different target architectures. Then we propose a new paradigm FirePower\footnote{It is open-sourced at https://github.com/hkust-zhiyao/FirePower} that targets the few-shot learning scenario for new target architectures by different users. 
    \item To the best of our knowledge, FirePower is the first data-driven architecture-level power model that explores cross-architecture design knowledge. This exploration contributes to the understanding of the gap among architectures. This framework is fully automated, without requiring additional designer knowledge about any target architecture.  
    \item FirePower proposes multiple new policies to use cross-archi-tecture knowledge. 1) It develops models at component level, and components are defined in power-friendly manner. 2) It supports different generalization strategies for different components. 3) It formulates generalizable and architecture-specific design knowledge into two separate models. 
    %\item FirePower also proposes a confidence-dependent knowledge generalization, utilizing the importance obtained from the foundation for feature selection.
    \item To detect the risk of huge differences between known and target architectures, FirePower supports the evaluation of the generalization quality for any target architecture. This helps provide the applicable scope of the method. 
    \item We evaluate FirePower using two widely adopted open-source RISC-V CPU designs: BOOM and XiangShan. It demonstrates that FirePower can achieve a low MAPE of 5.8\% and a high correlation $R$ of 0.98 on average only using two configurations of the target architecture. It achieves 8.8\% lower MAPE and 0.03 higher $R$ compared with directly training McPAT-Calib on configurations of the target architecture.
\end{itemize}

%\vspace{-.05in}
\section{Related Work}
%\vspace{-.03in}

Standard power estimation flow includes RTL implementation, logic synthesis, RTL simulation, and power simulation~\cite{powerpro, ptpx}. In recent years, design-specific ML power models have been proposed~\cite{kim2019simmani, zhou2019primal, xie2021apollo, kumar2019learning, kumar2022machine, lu2024unleashing, fang2023masterrtl, fang2024transferable, xie2022deep, peng2023prophet} for RTL stage using RTL signals as input. However, they still require RTL implementation, and a new model needs to be developed from scratch for each design. For pre-RTL stage, an accurate architecture-level power model is greatly needed.

Architecture-level power modeling takes architecture-level hardware parameters (denoted as $H$) and event statistics (denoted as $E$) as input to calculate power. Hardware parameters are the parameters used to describe the CPU configuration, such as $FetchWidth$ and $DCacheWay$. Event statistics are the information generated by running the workloads on the architecture-level performance simulator, such as the number of cache hits and branch instructions. 

Traditional analytical architecture-level power models like McPAT~\cite{li2009mcpat} calculate energy consumption for each event based on hardware parameters, then divide the accumulation of them by the execution time to calculate power. Wattch~\cite{brooks2000wattch} also adopts a similar method. It calculates the power of each cycle by accumulating the energy consumption for each event in this cycle and then dividing it by the time of a cycle. However, because of the discrepancy among the architecture-level performance simulator, the power model, and the actual CPU design, these analytical models are often inaccurate.

\label{panda}

To deal with the inaccuracy of the analytical power model, data-driven ML power models have been proposed in recent years~\cite{zhai2022mcpat,zhai2023microarchitecture,zhang2023panda}. One of the representative data-driven architecture-level power models is McPAT-Calib~\cite{zhai2022mcpat}. McPAT-Calib trains an ML model to calibrate McPAT output towards power labels. Denoting the output of McPAT as $M$, the McPAT-Calib can be formulated below.
%\vspace{-.05in}
\begin{equation*}
    P = \boldsymbol{F_{ml}} ({H,E,M})  \nonumber
        %\vspace{-.05in}
\end{equation*}
This ML model tries to capture the mapping from these features to the power, such mapping is obviously different for different architectures. In this case, it needs to be retrained from scratch when applied to a new architecture. 
The work of~\cite{zhai2023microarchitecture} uses the same formulation but performs transfer learning to a new domain of the same architecture design, where a domain means configurations with the same $DecodeWidth$. Therefore, it~\cite{zhai2023microarchitecture} still doesn't support cross-architecture power modeling. The PANDA~\cite{zhang2023panda} proposes human-crafted resource functions to achieve few-shot learning where the known configuration is limited. However, designing the resource function requires significant engineering expertise for each target architecture. This is not an automated solution.

%\vspace{-.05in}
\section{Problem Formulation}
\label{pf}
%\vspace{-.03in}
Here we introduce problem formulation. FirePower developer starts with a known architecture. The design space of the known architecture is already well-explored by the developer so there are sufficient known configurations. The power simulation of these configurations has also been performed with multiple workloads. The collection of the dataset is denoted as $\mathcal{D}_{known}$.

When applying FirePower, there are multiple in-progress projects that need to build power models for their own target architectures at a low cost. For these ongoing design projects, there are only a few available configurations of target architectures. The dataset with a few configurations of target architecture and corresponding power labels is denoted as $\mathcal{D}_{target}$. In experiments, we test three scenarios when there are 2, 3, or 4 available configurations in $\mathcal{D}_{target}$.

Our goal is to facilitate power modeling of these in-progress projects based on limited data of their target architectures $\mathcal{D}_{target}$. FirePower achieves this in two phases: knowledge extraction and application. In the knowledge extraction phase, the developer extracts generalizable knowledge based on known architecture $\mathcal{D}_{known}$. In the application phase, the knowledge is generalized to help develop power models for target architectures based on $\mathcal{D}_{target}$.

%These projects conduct the few-shot transfer. It utilizes the generalizable knowledge extracted by the developer to build their power models for their target architectures based on their limited data. To decide on accepting or rejecting the generalized power model, the generalization evaluation process is provided to estimate the generalization's quality using their training data.

%\vspace{-.05in}
\section{Methodology}
%\vspace{-.03in}
\subsection{Power Model Overview}
\label{powermodel}
%\vspace{-.03in}
The general FirePower solution is based on two basic insights. 

\textbf{Insight 1.} Instead of directly modeling total power, architecture-level power should be modeled for each \emph{power-friendly} component. Specifically, the whole design will be partitioned into multiple common components for power modeling purposes. Individual power models will be developed for each component, and total power is a summation of all component power. This brings multiple benefits: 1) compared with total power, individual components altogether provide more power labels; 2) designers have the flexibility to define components in a power-friendly manner, which is introduced in Section~\ref{comp-def}; and 3) smaller common components are affected by very few hardware parameters, thus component power model tends to be simple thus more general. It is further discussed in Insight 2.

\textbf{Insight 2.} When developing data-driven architecture-level power models, we observe that some knowledge tends to be more general, while others are more architecture-specific. Here more general knowledge refers to correlation patterns between basic component \emph{hardware parameters} $H$ and component \emph{hardware scale}, which describes the overall amount of logic (e.g., number of logic gates) in the component. One important reason is, the number of hardware parameters of each component is very limited, ranging from 1 to 4 in our experiment. Patterns based on fewer hardware parameters tend to be simpler and thus less ``overfit'' to a specific architecture. In contrast, knowledge related to event statistics $E$ involves not only complex event activities but also the interaction between events and hardware scale. This makes patterns related to event statistics relatively complex. We thus set this part architecture-specific.

Inspired by two aforementioned insights, FirePower chooses com-ponent-level power modeling, with each component's power decoupled into generalizable and architecture-specific parts. With hardware parameters of the $i$-th component as $H_i$ and event statistics as $E_i$, the FirePower power model can be formulated below.
%\vspace{-.06in}
\begin{equation*}
    P^i = \boldsymbol{F_{hw}}^i (H_i) * \boldsymbol{F_{event}}^i (H_i, E_i) \nonumber \label{eq:panda}
    %\vspace{-.05in}
\end{equation*}
The $\boldsymbol{F_{hw}}^i$ denotes the hardware model of component $i$, which learns the basic correlation between hardware scale and hardware parameters. The $\boldsymbol{F_{event}}^i$ denotes the event model, an ML model to capture more complex correlations related to event statistics.
%learning more complex event-statistics-related knowledge. 
As for why multiplication is used to associate these two models, power consumption $P^i$ is roughly proportional to hardware scale (e.g., number of logic gates) in $\boldsymbol{F_{hw}}^i$, assuming a constant average toggle rate. The event-related knowledge captured by $\boldsymbol{F_{event}}^i$ partially reflects the toggle rate of the real workload. The toggle rate is also proportional to power. Multiplication is the simplest operator to capture the linear relationship between these two models and power.

The remainder of this section will introduce the FirePower methodology in detail. Section~\ref{comp-def} describes our proposed power-friendly component definition for power modeling at the component level. Section~\ref{phase1} and Section~\ref{phase2} will introduce the knowledge extraction (Phase 1) and application (Phase 2) of FirePower, respectively. 

\begin{figure}[!t]
\centering
%\vspace{-7mm}
\includegraphics[width=0.35\textwidth]{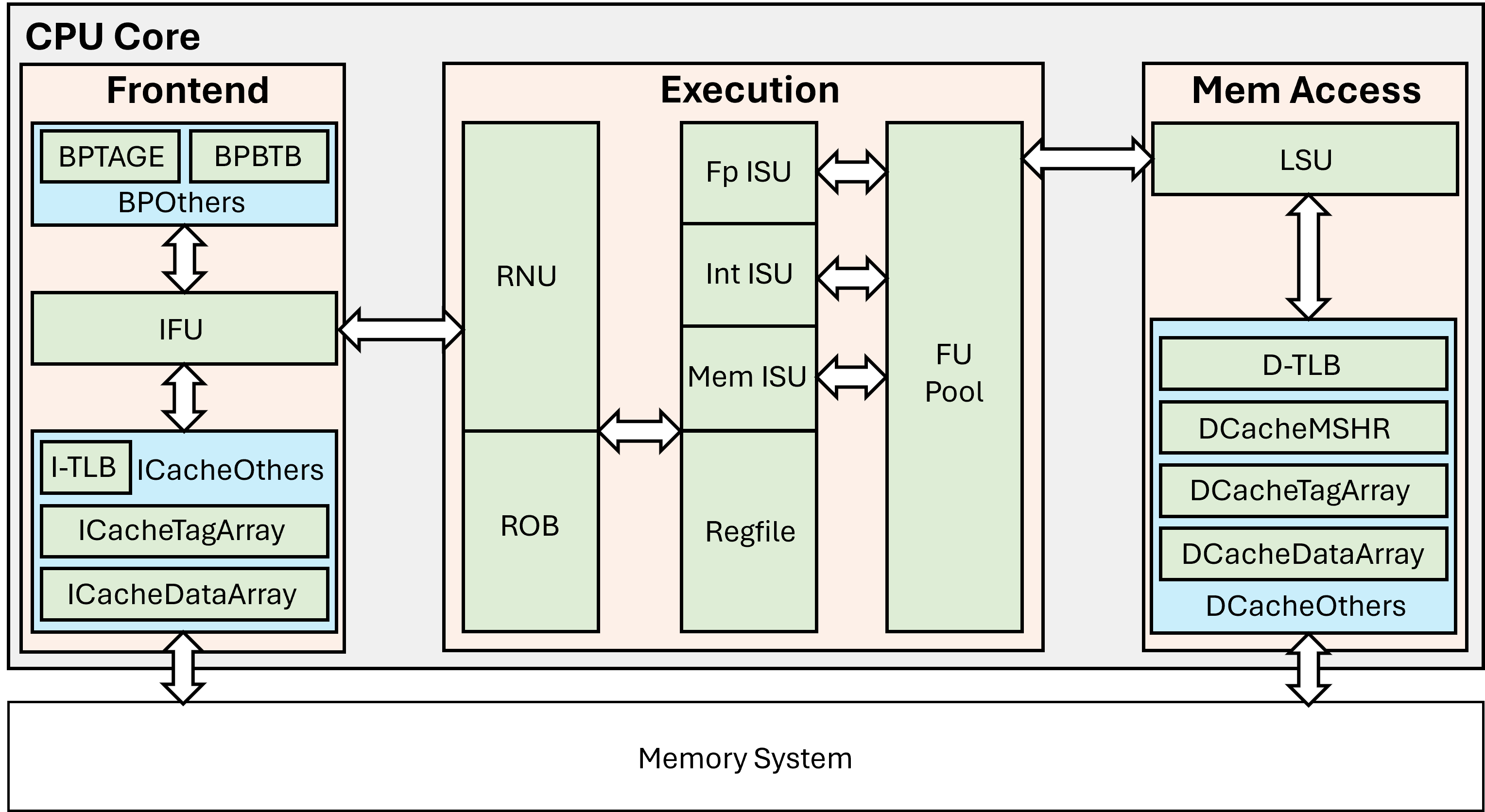}
\vspace{-3mm}
\caption{The illustration of our power-friendly component definition for the out-of-order CPU core.} 
\vspace{-3mm}
\label{compfig}
\end{figure}

%\vspace{-.1in}
\subsection{Power-Friendly Component Definition}
\label{comp-def}
\vspace{-.03in}
\begin{table}[!t]
      \centering
      %\vspace{-4mm}
      \renewcommand{\arraystretch}{1.1}
      \resizebox{0.36\textwidth}{!}{
        \begin{tabular}{ |c|c|c| } 
        \hline
        \multirow{2}{*}{Component $i$} &  Hardware Parameters of  &  Important  \\
                                       &  Each Component $H_i$   &  Parameter  \\
        \hline
         \hline
BPTAGE &  FetchWidth, BranchCount  & FetchWidth  \\
         \cline{1-3}
BPBTB &  FetchWidth, BranchCount  & FetchWidth   \\
         \cline{1-3}
BPOthers &  FetchWidth, BranchCount  &  FetchWidth  \\
         \cline{1-3}
\multirow{2}{*}{IFU} &  FetchWidth, DecodeWidth,   &  \multirow{2}{*}{--}    \\
         &  FetchBufferEntry, ICacheFetchBytes  &   \\ 
         \cline{1-3}
I-TLB &  ICacheTLBEntry  & --  \\
         \cline{1-3}
%I-Cache &  ICacheWay, ICacheFetchBytes  & icache.overallAccesses, icache.overallMisses, icache.ReadReq.mshrHits, icache.ReadReq.mshrMisses, icache.tagAccesses & \\ 
%\hline
         %& & ReadReq.mshrMisses, tagAccesses   \\
ICacheTagArray &  ICacheWay, ICacheFetchBytes & DCache/ICacheWay  \\  
\cline{1-3}
ICacheDataArray &  ICacheWay, ICacheFetchBytes & FetchWidth  \\  
\cline{1-3}
ICacheOthers &  ICacheWay, ICacheFetchBytes & --  \\  
         \hline
         \hline

RNU & DecodeWidth  & DecodeWidth  \\
         \cline{1-3}
ROB &  DecodeWidth, RobEntry  & --  \\
         \cline{1-3}
FP ISU &  DecodeWidth, FpIssueWidth   & --   \\
        \cline{1-3}
Int ISU &  DecodeWidth, IntIssueWidth,   &  DecodeWidth  \\
        \cline{1-3}
Mem ISU &  DecodeWidth, MemIssueWidth   & --  \\
        \cline{1-3}
Regfile &  DecodeWidth, IntPhyRegister, FpPhyRegister  & --   \\
        % & & fpRegfileWrites, functionCalls\\
         \cline{1-3}
FU Pool &  Mem/FpIssueWidth, IntIssueWidth  & Mem/FpIssueWidth   \\
        \hline
        \hline

LSU &  LDQEntry, STQEntry, MemIssueWidth  & --  \\
         \cline{1-3} 
D-TLB &  DCacheTLBEntry  & DTLBEntry \\
         \cline{1-3}
\multirow{2}{*}{DCacheTagArray} &  DCacheWay, DCacheTLBEntry,  & \multirow{2}{*}{--}   \\
         & MemIssueWidth  &   \\
         \cline{1-3} 
\multirow{2}{*}{DCacheDataArray} &  DCacheWay, DCacheTLBEntry,  & \multirow{2}{*}{--}   \\
         & MemIssueWidth  &  \\
         \cline{1-3} 
DCacheMSHR &  MSHREntry  & MSHREntry   \\
         \cline{1-3} 
\multirow{2}{*}{DCacheOthers} &  DCacheWay, DCacheTLBEntry,  &  \multirow{2}{*}{--}  \\
         & MSHREntry, MemIssueWidth  &  \\
        \hline
        \hline
         Other Logic &  All  & -- \\
        \hline
        \end{tabular}
        }
        \caption{Our identified architecture-level hardware parameters and the important parameter detected for Retraining.}
        \label{tbl:config_event}
        \vspace{-8mm}
\end{table}

To facilitate component-level data-driven power modeling, we propose a power-friendly component definition for out-of-order CPU. The component definition has two targets: being common and fine-grained. 1) To generalize power model to different microarchitecture designs, the component definition should be common, where each component can be found in different out-of-order CPUs. %\yao{[general is not clear here, what does it mean?]} 
2) To facilitate per-component power modeling, component definition should be fine-grained so that the circuit in the same component should correlate with similar hardware parameters and event statistics. %\yao{requirement -> metrics} 

To meet the two targets above, we propose a power-friendly component definition. %, which is illustrated in Fig.~\ref{compfig}. 
The components are in three main parts: Frontend, Execution, and Mem Access. Details are introduced below.
\begin{itemize}
\item The Frontend includes 8 components: TAGE in branch predictor (BPTAGE), BTB in branch predictor (BPBTB), others in branch predictor (BPOthers), instruction fetch unit (IFU), instruction translation lookup buffer (I-TLB), instruction cache tag array (ICacheTagArray), instruction cache data array (ICacheDataArray), and others in instruction cache (ICacheOthers).
\item The Execution consists of 7 components: renaming unit (RNU), reorder buffer (ROB), issue unit of float point instruction (Fp ISU), issue unit of integer instruction (Int ISU), issue unit of memory access instruction (Mem ISU), register file (Regfile), and function unit pool (FU Pool).
\item The Mem Access has 6 components: load-store unit (LSU), data translation lookup buffer (D-TLB), miss status handle register (DCacheMSHR), data cache tag array (DCacheTagArray), data cache data array (DCacheDataArray), and others in data cache (DCacheOthers).
\item Other circuits not covered by the above components are referred to as a new component named Other Logic. 
\end{itemize}
%The associated hardware parameters of each component are shown in Table.~\ref{tbl:config_event}, where the associated event statistics are not listed because of the page limitation.
Table~\ref{tbl:config_event} shows the associated hardware parameters of each component, where event statistics are not listed because of page limitation.

\begin{figure}[!t]
%\vspace{-9mm}
\centering
%\hspace{-6mm}
\subfigure[DCacheDataArray]{
    \centering
    \includegraphics[width=0.20\textwidth]{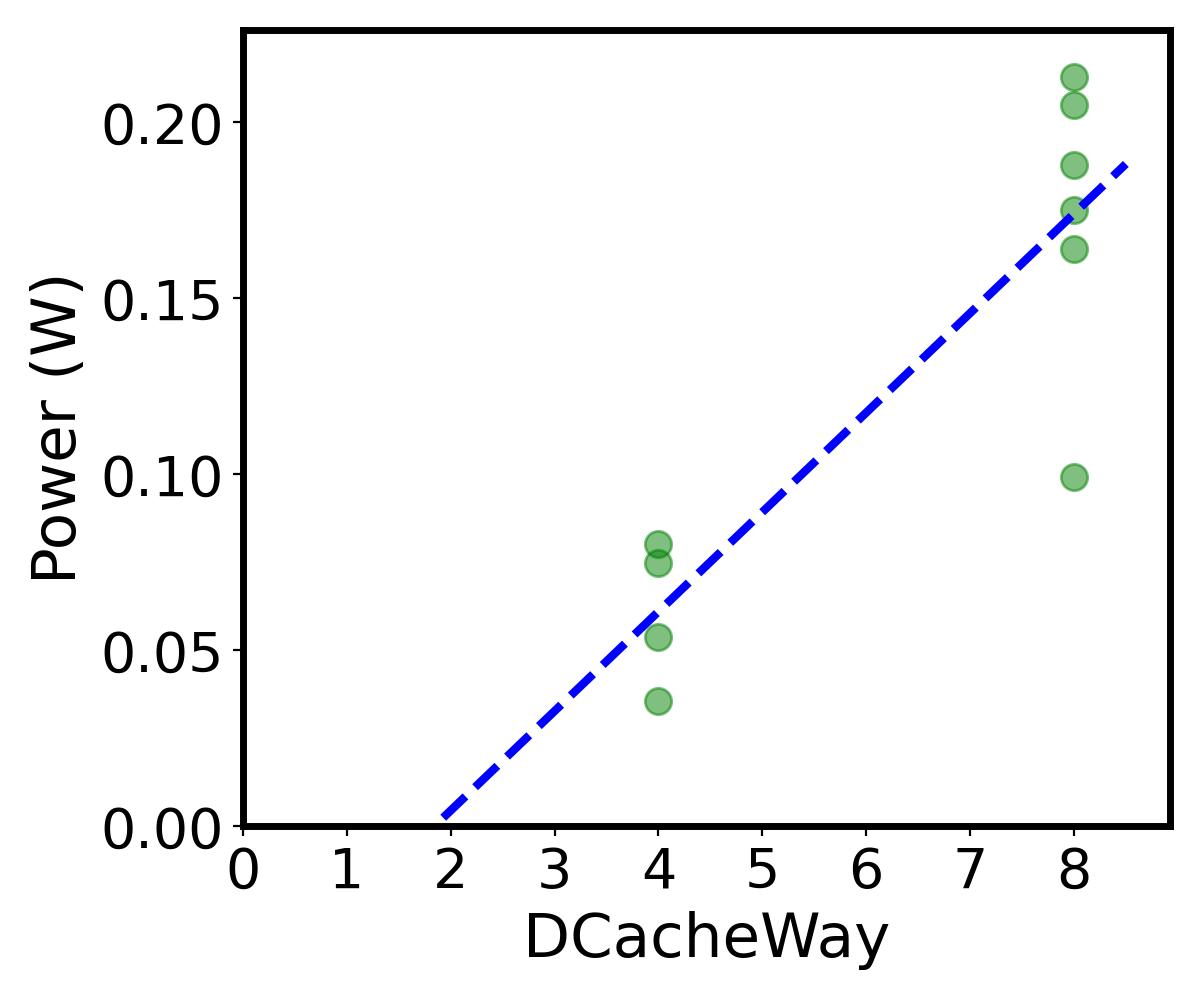}
}
%\hspace{1mm}
\subfigure[DCacheMSHR]{
    \centering
    \includegraphics[width=0.20\textwidth]{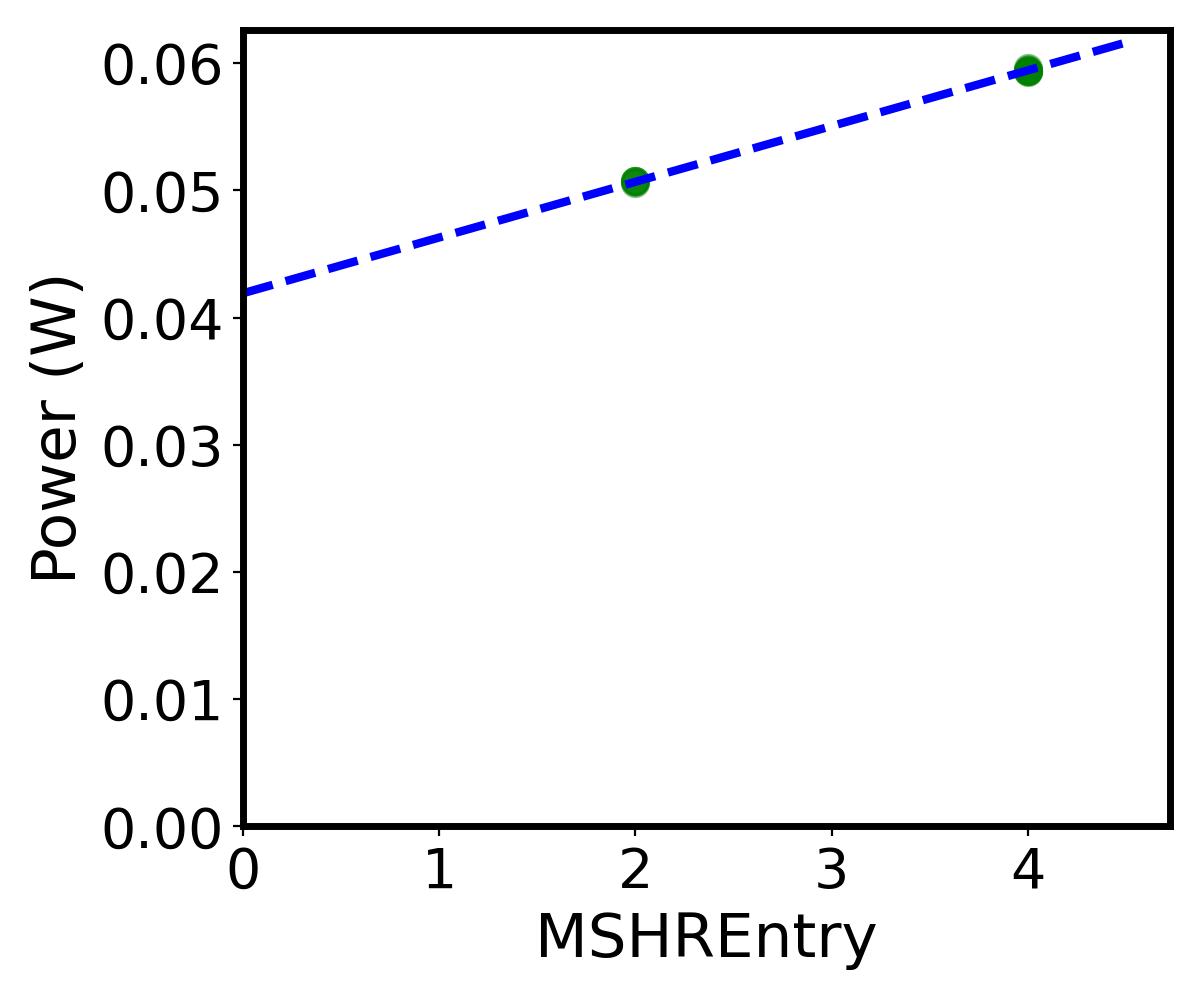}
}
\vspace{-5mm}
\caption{Correlation between power and the most related hardware parameter. The two components correlate with different hardware parameters. DCacheDataArray correlates with $DCacheWay$, DCacheMSHR correlates with $MSHREntry$.} %\yao{Too blurred. Replace the Fig. What does this Fig demonstrate? Write down your key point here. Why their scaling trends are different? Different slope? Then mark the slope in both figures.}}
\vspace{-5mm}
\label{scaling}
\end{figure}

%\yao{extend this example}

%%\yao{``break CPU down'' is incorrect, it should be ``decompose into''}    

%\yao{[Start with none of prior work use components except PANDA. Then mention the difference from PANDA]} 
%  decomposes the CPU into components, it is

No existing architecture-level ML power modeling works built component-level models except PANDA~\cite{zhang2023panda}. The component power model in PANDA~\cite{zhang2023panda} is based on default component partitioning without considering power. For example, in~\cite{zhang2023panda}, the DCache is a whole component, but within this DCache, we can find that the DCacheDataArray highly correlates with a hardware parameter $DCacheWay$, while DCacheMSHR correlates with another hardware parameter $MSHREntry$. The correlation is illustrated in Fig.~\ref{scaling} using the XiangShan CPU~\cite{xu2022towards}, showing their correlations with different hardware parameters. They are thus separated and processed with different power models in FirePower to capture clearer patterns. It conforms to the aforementioned fine-granularity target.

%\vspace{-.1in}
\subsection{Phase 1: Knowledge Extraction}
\label{phase1}
%\vspace{-.03in}
In phase 1 of FirePower, developers perform knowledge extraction, extracting the generalizable knowledge based on sufficient data of an already known architecture. This process only needs to be performed once by solution developers, as shown in the left of Fig.~\ref{method}. Based on the known architecture, generalizable knowledge is extracted for each component, including two types of information: (1) hardware model \textbf{$F_{hw}^i$} built on the known architecture, (2) the importance of the hardware parameters of this component. 

%\yao{[summarize why importance matters]}

%\vspace{-1mm}
\subsubsection{Hardware Model}

The hardware model $\boldsymbol{F_{hw}}^i$  learns the relationship between hardware scale and hardware parameters, as introduced in the overview. To represent the hardware scale as labels, we calculate the average power across all workloads. This average power label reflects the general power characteristics across workloads. The input features, as summarized in Table~\ref{tbl:config_event}, are hardware parameters $H_i$ of each component. The ML model we use is the XGBoost~\cite{chen2016xgboost}, which is one of the most widely adopted regressors. It is a decision-tree-based ensemble learning algorithm using an ensemble of weak prediction models for regression.

%\vspace{-1mm}
\subsubsection{Parameter Importance}

In addition to the hardware model, we will further evaluate the \emph{importance} of hardware parameters $H_i$ for each component. Such importance reflects the impact of each hardware parameter on the component power. The distribution of parameter importance will help us analyze the power correlation pattern for each component: The key idea is to evaluate whether there is a dominating hardware parameter for each component. It will affect how to generalize the knowledge in phase 2. More details about the usage of parameter importance are discussed in Sec.~\ref{kg}. 

\begin{figure}[!t]
\centering
%\hspace{-4mm}
%\vspace{-4mm}
\includegraphics[width=0.42\textwidth]{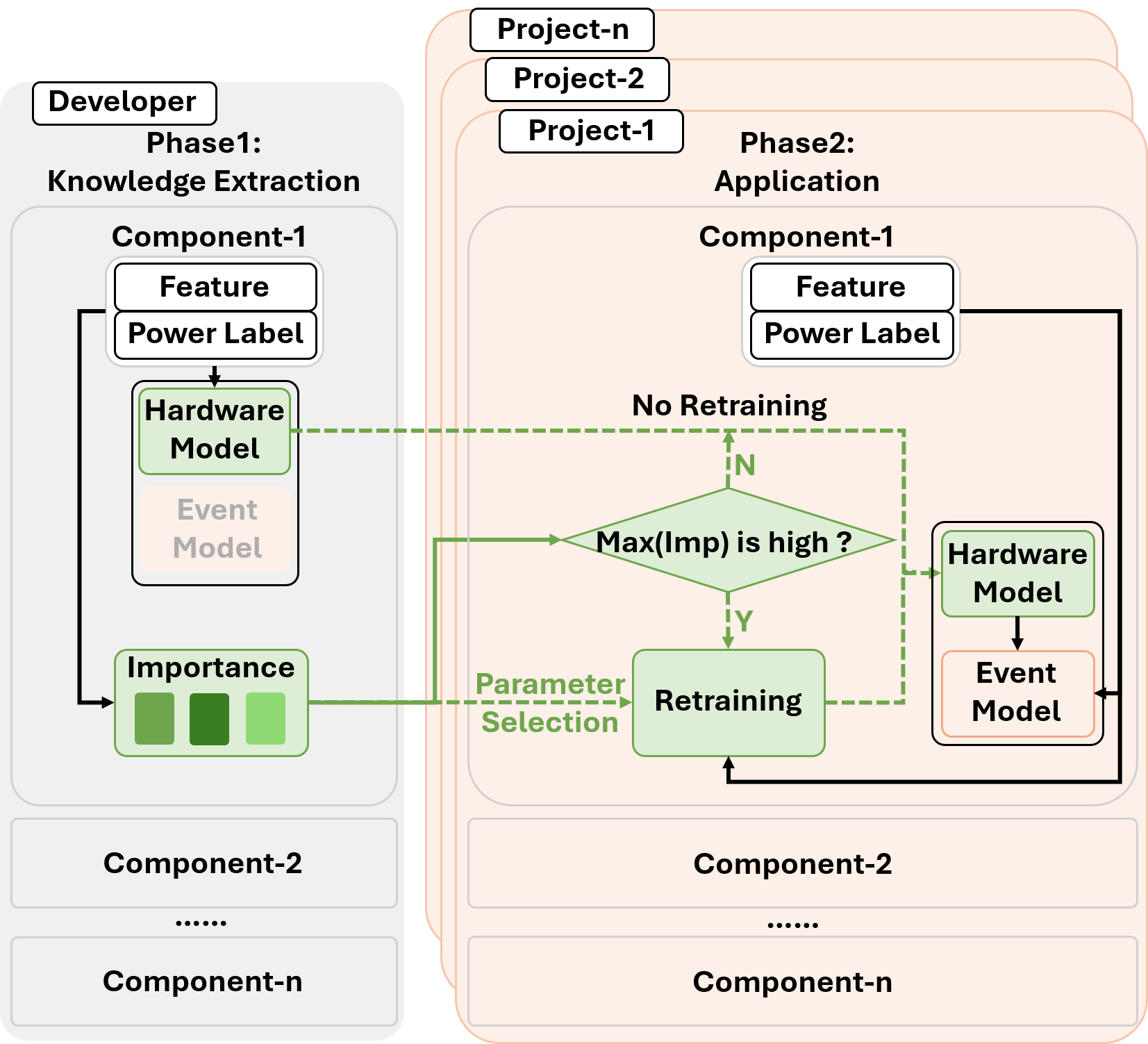}
\vspace{-3mm}
\caption{The FirePower framework with two phases. Knowledge extraction in phase 1 extracts hardware model and parameter importance from a known architecture as general knowledge. Application in phase 2 adopts two knowledge generalization strategies, Retraining and No Retraining, depending on the parameter importance distribution.}
\vspace{-4mm}
\label{method}
\end{figure}

The hardware parameter importance is calculated based on the hardware model. Such a tree-model-based evaluation calculates feature importance based on impurity decreases contributed by each feature (e.g., parameter)~\cite{breiman2001random}. Importance evaluation is not limited to tree models, there are also some methods, such as SHAP~\cite{lundberg2017unified}, to evaluate parameter importance for arbitrary ML models.

%\vspace{-.1in}
\subsection{Phase 2: Application}
\label{phase2}
%\vspace{-.03in}
The application as phase 2 of FirePower is shown in the right of Fig.~\ref{method}. Compared to knowledge extraction, which is performed only once in total by the developer, the application phase can be applied many times to different projects and target architectures. In each project, phase 2 utilizes the extracted generalizable knowledge and the limited data of the target architecture to build the power model for the target architecture. Specifically, the application phase performs three steps: (1) apply the generalizable knowledge to build the hardware model $\boldsymbol{F_{hw}}^i$, (2) train the event model $\boldsymbol{F_{event}^i}$, and (3) evaluate the generalization to estimate the generalization's quality. 

%\vspace{-2mm}
\subsubsection{Hardware Model}
\label{kg}

For each component, we support two knowledge generalization strategies, named Retraining and No Retraining, as illustrated in Fig.~\ref{method}.

\textbf{Strategy 1:} No Retraining is straightforward, it directly adopts hardware model $\boldsymbol{F_{hw}}^i$ trained on known architecture dataset $\mathcal{D}_{known}$ for target one. 
Directly applying hardware model may not accurately estimate average power due to differences in hardware scale of different architectures. The rationale behind No Retraining is hardware model primarily captures the correlation rather than the absolute average power value. The ratio between known and target architecture will be captured by additional event model $\boldsymbol{F_{event}^i}$.

\textbf{Strategy 2:} For Retraining, it trains a brand new hardware model using limited available configurations $\mathcal{D}_{target}$ of the target architecture. Such a retraining faces a trade-off. On the one hand, available configurations directly from target architecture naturally help transfer. On the other hand, since available configurations in $\mathcal{D}_{target}$ are limited, the model can easily overfit. To avoid overfitting, the key idea behind retraining is to maximally simplify new hardware model $\boldsymbol{F_{hw}}^i$. This retrained hardware model $\boldsymbol{F_{hw}}^i$ will be a linear model based on one most important parameter from $H_i$. Retraining policy is more suitable for components with relatively simple power characteristics with one dominating important parameter.

\textbf{Strategy selection:} For each component, we select the most appropriate strategy (Retraining vs. No Retraining) based on parameter importance distribution. Such parameter importance is also the knowledge extracted from known architecture in phase 1. When only one hardware parameter has a dominating importance, it indicates the overall power correlation is simple. On the contrary, if the distribution is more uniform, it means this component is relatively complex. Therefore, if maximum parameter importance exceeds a threshold\footnote{We set the threshold to 0.95 in the experiment, with the sum of all hardware parameters $H_i$ in each component normalized to 1.}, Retraining will be adopted. Otherwise, No Retraining is selected. The strategy selection result based on BOOM CPU architecture is listed in Table~\ref{tbl:config_event}, where the important parameter used for Retraining is listed, and the ``--" means No Retraining.

% (e.g., 0.95)
%\vspace{-2mm}
\subsubsection{Event Model}

The event model $\boldsymbol{F_{event}^i}$ will mainly capture the more complex correlation related to event statistics, which is highly architecture-specific. To train the event model for target architectures, for each component, we take both hardware parameters $H_i$ and event statistics $E_i$ of each component as features. The event model's training label is $P_i/\boldsymbol{F_{H}^i}$, which is the ratio between the component power label in $\mathcal{D}_{target}$ and the hardware model $\boldsymbol{F_{H}^i}$. The adopted ML model is also XGBoost~\cite{chen2016xgboost}.

%\vspace{-.1in}
\subsection{Generalization Quality Evaluation}
\label{ge}
%\vspace{-.03in}
The effectiveness of knowledge generalization can be compromised when there is a significant difference between the target architecture and the known architecture. Hence, it is crucial to evaluate the generalization's quality to help determine whether to accept generalized power model or resort to time-consuming traditional power modeling paradigm. Such quality evaluation helps indicate \emph{applicable scope} of FirePower based on the known architecture.

For such generalization evaluation on each component, we compare average power labels of each target configuration from $\mathcal{D}_{target}$ with predictions of hardware model $\boldsymbol{F_{hw}^i}$ from phase 1, without considering retraining. Such $\boldsymbol{F_{hw}^i}$ reflects known architecture $\mathcal{D}_{known}$, and thus the comparison indicates the difference between known and target architecture. When comparing, we adjust prediction of hardware model by multiplying it with an ideal scaling factor, since hardware model $\boldsymbol{F_{hw}^i}$ only captures the trend and detailed ratios are left for event models  $\boldsymbol{F_{event}^i}$, as discussed in Sec.~\ref{kg}.

\section{Experiment Setup}

\begin{table*}[!t]
\centering
      %\vspace{-.35in}
    %  \renewcommand{\arraystretch}{1.05}
      \resizebox{0.8\textwidth}{!}{
        \begin{tabular}{ |c||c c c c c c c c c c c c c c c||c c c c c c c c c c| } 
\hline
Hardware Parameter  & B1 & B2 & B3 & B4 & B5 & B6 & B7 & B8 & B9 & B10 & B11 & B12 & B13 & B14 & B15 & X1 & X2 & X3 & X4 & X5 & X6 & X7 & X8 & X9 & X10\\
\hline
\hline
FetchWidth & 4 & 4 & 4 & 4 & 4 & 8 & 8 & 8 & 8 & 8 & 8 & 8 & 8 & 8 & 8 &      4 & 4 & 4 & 4 & 4 & 8 & 8 & 8 & 8 & 8\\
\hline
DecodeWidth & 1 & 1 & 1 & 2 & 2 & 2 & 3 & 3 & 3 & 4 & 4 & 4 & 5 & 5 & 5 &     2 & 2 & 2 & 3 & 3 & 3 & 4 & 4 & 4 & 5\\
\hline
FetchBufferEntry & 5 & 8 & 16 & 8 & 16 & 24 & 18 & 24 & 30 & 24 & 32 & 40 & 30 & 35 & 40 &    8 & 16 & 24 & 16 & 24 & 24 & 24 & 32 & 32 & 24\\
\hline
RobEntry & 16 & 32 & 48 & 64 & 64 & 80 & 81 & 96 & 114 & 112 & 128 & 136 & 125 & 130 & 140 &     16 & 32 & 48 & 64 & 64 & 80 & 81 & 96 & 114 & 112\\
\hline
IntPhyRegister & 36 & 53 & 68 & 64 & 80 & 88 & 88 & 110 & 112 & 108 & 128 & 136 & 108 & 128 & 140 &     36 & 53 & 68 & 64 & 80 & 88 & 88 & 110 & 112 & 108\\
\hline
FpPhyRegister & 36 & 48 & 56 & 56 & 64 & 72 & 88 & 96 & 112 & 108 & 128 & 136 & 108 & 128 & 140 &     36 & 53 & 68 & 64 & 80 & 88 & 88 & 110 & 112 & 108\\
\hline
LDQ/STQEntry & 4 & 8 & 16 & 12 & 16 & 20 & 16 & 24 & 32 & 24 & 32 & 36 & 24 & 32 & 36 &     16 & 20 & 24 & 20 & 24 & 28 & 24 & 32 & 40 & 32\\
\hline
BranchCount & 6 & 8 & 10 & 10 & 12 & 14 & 14 & 16 & 16 & 18 & 20 & 20 & 18 & 20 & 20 &     7 & 7 & 7 & 7 & 7 & 7 & 7 & 7 & 7 & 7\\
\hline
Mem/FpIssueWidth & 1 & 1 & 1 & 1 & 1 & 1 & 1 & 1 & 2 & 1 & 2 & 2 & 2 & 2 & 2 &     2 & 2 & 2 & 2 & 2 & 2 & 2 & 2 & 2 & 2\\
\hline
IntIssueWidth & 1 & 1 & 1 & 1 & 2 & 2 & 2 & 3 & 3 & 4 & 4 & 4 & 5 & 5 & 5 &     2 & 2 & 2 & 2 & 4 & 4 & 4 & 6 & 6 & 6\\
\hline
DCache/ICacheWay & 2 & 4 & 8 & 4 & 4 & 8 & 8 & 8 & 8 & 8 & 8 & 8 & 8 & 8 & 8 &     4 & 4 & 8 & 4 & 4 & 8 & 8 & 8 & 8 & 8\\
\hline
DTLBEntry & 8 & 8 & 16 & 8 & 8 & 16 & 16 & 16 & 32 & 32 & 32 & 32 & 32 & 32 & 32 &     8 & 8 & 16 & 8 & 8 & 16 & 16 & 16 & 32 & 32\\
\hline
MSHREntry & 2 & 2 & 4 & 2 & 2 & 4 & 4 & 4 & 4 & 4 & 4 & 8 & 8 & 8 & 8 &     2 & 2 & 4 & 2 & 2 & 4 & 4 & 4 & 4 & 4\\
\hline
ICacheFetchBytes & 2 & 2 & 2 & 2 & 2 & 4 & 4 & 4 & 4 & 4 & 4 & 4 & 4 & 4 & 4 &     2 & 2 & 2 & 2 & 2 & 2 & 2 & 2 & 2 & 2\\
         \hline
        \end{tabular}
        }
    %   \vspace{-.08in}
        %\vspace{-1mm}
        \caption{The CPU configurations used in our experiment. The B1-B15 denote the 15 configurations of BOOM, and the X1-X10 denote the 10 configurations of XiangShan.}
        \vspace{-9mm}
        \label{configtable}
\end{table*}

%\vspace{-.03in}
\subsection{RISC-V CPU Cores for Experiment}
%\vspace{-.02in}
In our experiment, we adopt two different RISC-V CPU cores as our experimented architectures. RISC-V~\cite{URL:riscv} is one of the most widely adopted open-source instruction set architecture. Nowadays, most open-source CPU design projects are based on RISC-V, the most representative two projects of which are BOOM~\cite{zhao2020sonicboom} and XiangShan~\cite{xu2022towards}. BOOM and XiangShan are both highly configurable, enabling us to generate different configurations for each architecture. They are both out-of-order CPU cores with similar major CPU components but there are also many differences. Although they both use RISC-V, the version is not the same, with RV64GC for BOOM and RV64GCBK for XiangShan. Besides, architectural designs for components are different. Taking the second-level branch predictor as an example, BOOM adopts traditional BTB for target prediction, while XiangShan adopts Fetch Target Buffer to replace BTB. %Each FTB entry forms a prediction block, and it predicts both the starting address of the next block and the endpoint of the current block. 
Considering the reasonable similarities and differences between BOOM and XiangShan, we evaluate our paradigm on them. 

When evaluating each method, we conduct multiple experiments with different known/target architecture settings and different numbers of configurations of target architecture. Because of limited accessible open-source RISC-V CPU architectures, there is only one target architecture. Known/target architecture settings include BOOM as known architecture and XiangShan as target architecture denoted as BOOM$\,\to\,$XS and vice versa denoted as XS$\,\to\,$BOOM. For different numbers of configurations of target architecture, in knowledge extraction (Phase 1), all known architecture configurations are used. For application (Phase 2), configurations of target architecture are used for knowledge generalization, and remaining ones are used for testing. We evaluate the accuracy with mean absolute percentage error (MAPE) and correlation coefficient $R$. 

Configurations adopted for BOOM and XiangShan are listed in Table~\ref{configtable}, covering different scales. There are 15 configurations for BOOM named B1 to B15 and 10 for XiangShan named X1 to X10. To reflect the scenario where only a few configurations of target architecture are accessible, the number of labeled configurations of target architecture is set as 4, 3, and 2 for three sets of experiments. %, which are randomly selected. 
%For an in-development project, configurations of different scales are implemented preferentially to better cover the design space. Based on this fact, we select the labeled configurations of target architecture  

%\vspace{-.05in}
\subsection{Design Implementation Flow}
\label{setup}
%\vspace{-.02in}
In our experiment, to collect the dataset, RTL code generation and RTL simulation of BOOM CPU~\cite{zhao2020sonicboom} is performed with Chipyard~\cite{amid2020chipyard} v1.8.1, and that of XiangShan CPU~\cite{xu2022towards} is performed with OpenXiangShan~\cite{URL:openxs}. %Using these two frameworks, we generate 15 configurations for BOOM named B1 to B15, and 10 configurations in XiangShan, named X1 to X10, which are both shown in Table~\ref{configtable}, covering design configurations with different scales.
For workload-driven power simulation to generate ground truth power, we used eight workloads in riscv-tests~\cite{URL:riscvtests} including dhrystone, median, multiply, qsort, rsort, towers, spmv, and vvadd. Minor modifications are made for adaptation on XiangShan.

The RTL simulation is performed with Synopsys VCS\textsuperscript{\textregistered}~\cite{vcs}. We performed logic synthesis and power simulation with Synopsis Design Compiler\textsuperscript{\textregistered}~\cite{design-compilier} and PrimePower~\cite{ptpx} respectively. Our VLSI flow is based on TSMC 40nm standard cell library and associated Memory Compiler for SRAM generation. For the microarchitecture simulation, we use gem5~\cite{binkert2011gem5} as performance simulator to generate event statistics. We also use McPAT~\cite{li2009mcpat} as power model to generate power estimation as some of features for McPAT-Calib.

%\vspace{-.05in}
\subsection{Summary of Baseline Methods}
%\vspace{-.02in}
We compare FirePower with the state-of-the-art architecture-level power model McPAT-Calib~\cite{zhai2022mcpat} as our baseline. The other work~\cite{zhang2023panda} is not included since it is not an automated method, requiring engineer-defined functions. Besides McPAT-Calib~\cite{zhai2022mcpat}, we further include four more ablation studies based on part of FirePower's policies. (1) The McPAT-Calib + Component. It builds ML models for each component, using the associated hardware parameters and event statistics as features and per-component power as labels. (2) The McPAT-Calib + Transfer Learning. It adopts the McPAT-Calib as the power modeling method, builds a model on known architecture as the source model, and then adopts one of the most widely adopted transfer learning algorithms, pseudo label~\cite{lee2013pseudo}, for knowledge generalization. In detail, for each testing data of the target architecture, we search for the nearest labeled data of the target architecture using the distance in feature space, where the label is $L$. We use the source model to make predictions for the testing data and its nearest sample, denoted as $p_t$ and $p_l$. The prediction on test data is $P_t = \frac{p_t}{p_l}L$. (3) The McPAT-Calib + Component + Transfer Learning. It combines (1) and (2), performing transfer learning for each component respectively. (4) FirePower without Retraining. It only adopts the No Retraining as the knowledge generalization strategy, without taking the parameter importance as the generalizable knowledge. For a fair comparison, for all baselines and our FirePower solution, we adopt the same XGBoost~\cite{chen2016xgboost}, which is the best ML model reported in McPAT-Calib~\cite{zhai2022mcpat}, with default hyperparameters, i.e. n\_estimator=100 and depth=3.

%\vspace{-.05in}
\section{Experimental Results}
%\vspace{-.02in}
\subsection{Power Modeling Accuracy}
%\vspace{-.02in}
\begin{figure}[!t]
%\vspace{-1mm}
%\hspace{-7mm}

\centering
\captionsetup[subfigure]{aboveskip=-1pt,belowskip=-1pt}
\subfigure[XS$\,\to\,$BOOM]{
    \centering
    \includegraphics[width=0.45\textwidth]{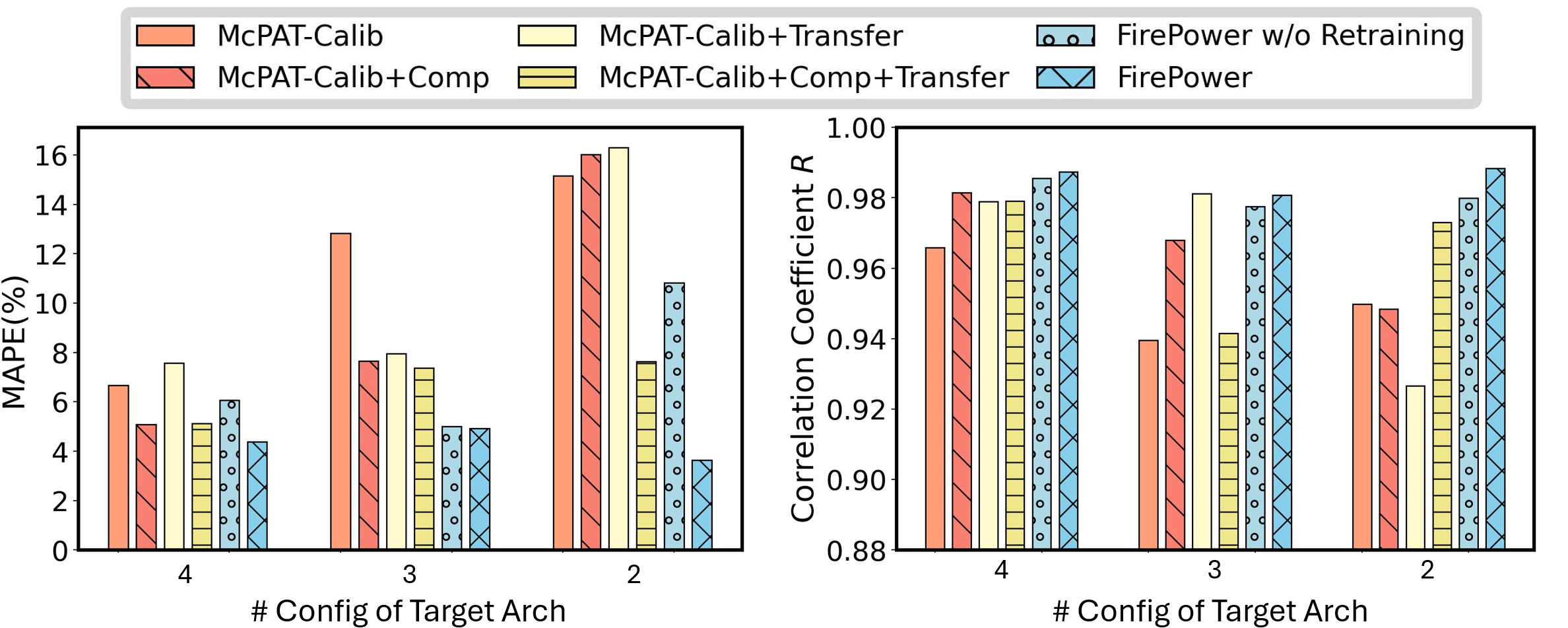}
}
\vspace{-3mm}

\subfigure[BOOM$\,\to\,$XS]{
    \centering
    \includegraphics[width=0.45\textwidth]{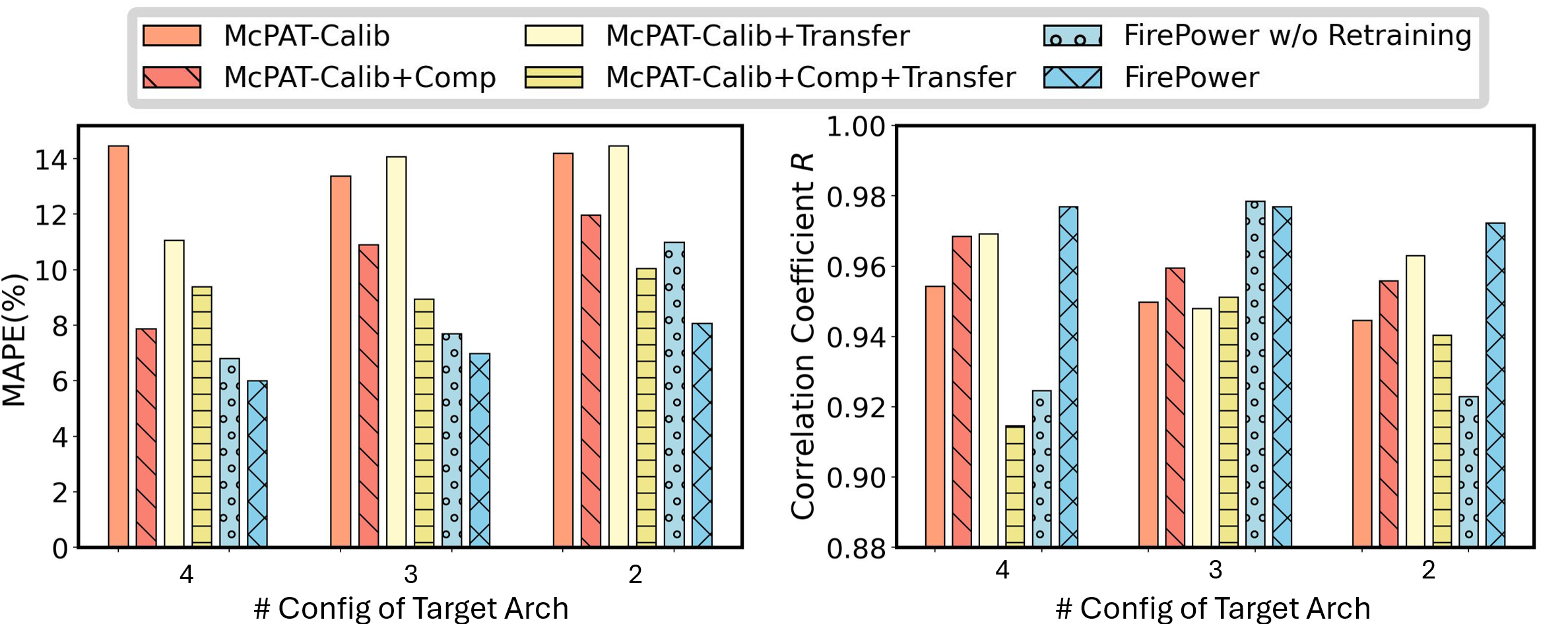}
}
\vspace{-4mm}
\caption{Summary of the comparison between FirePower and other methods under different numbers of configurations of target architecture. ``Comp'' stands for Component and ``Transfer'' stands for Transfer Learning.}
\vspace{-4mm}
\label{ablation}
\end{figure}

\begin{figure}[!t]
\centering
%\vspace{-6mm}

%\hspace{-6mm}
\subfigure[McPAT-Calib (BOOM)]{
    \centering
    \includegraphics[width=0.20\textwidth]{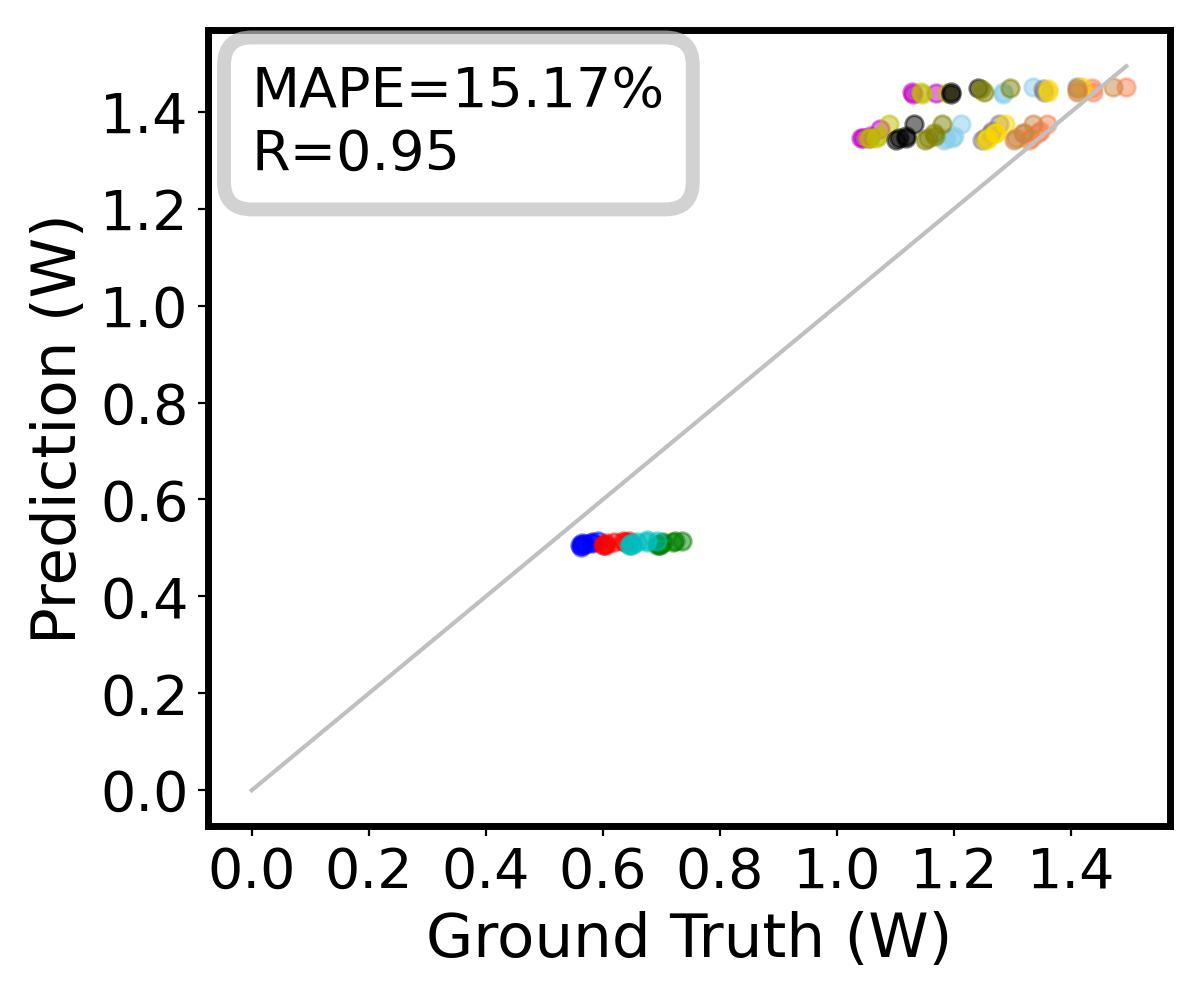}
    \label{mcpat_xs2boom2}
}
%\hspace{-2mm}
\subfigure[McPAT-Calib (XS)]{
    \centering
    \includegraphics[width=0.20\textwidth]{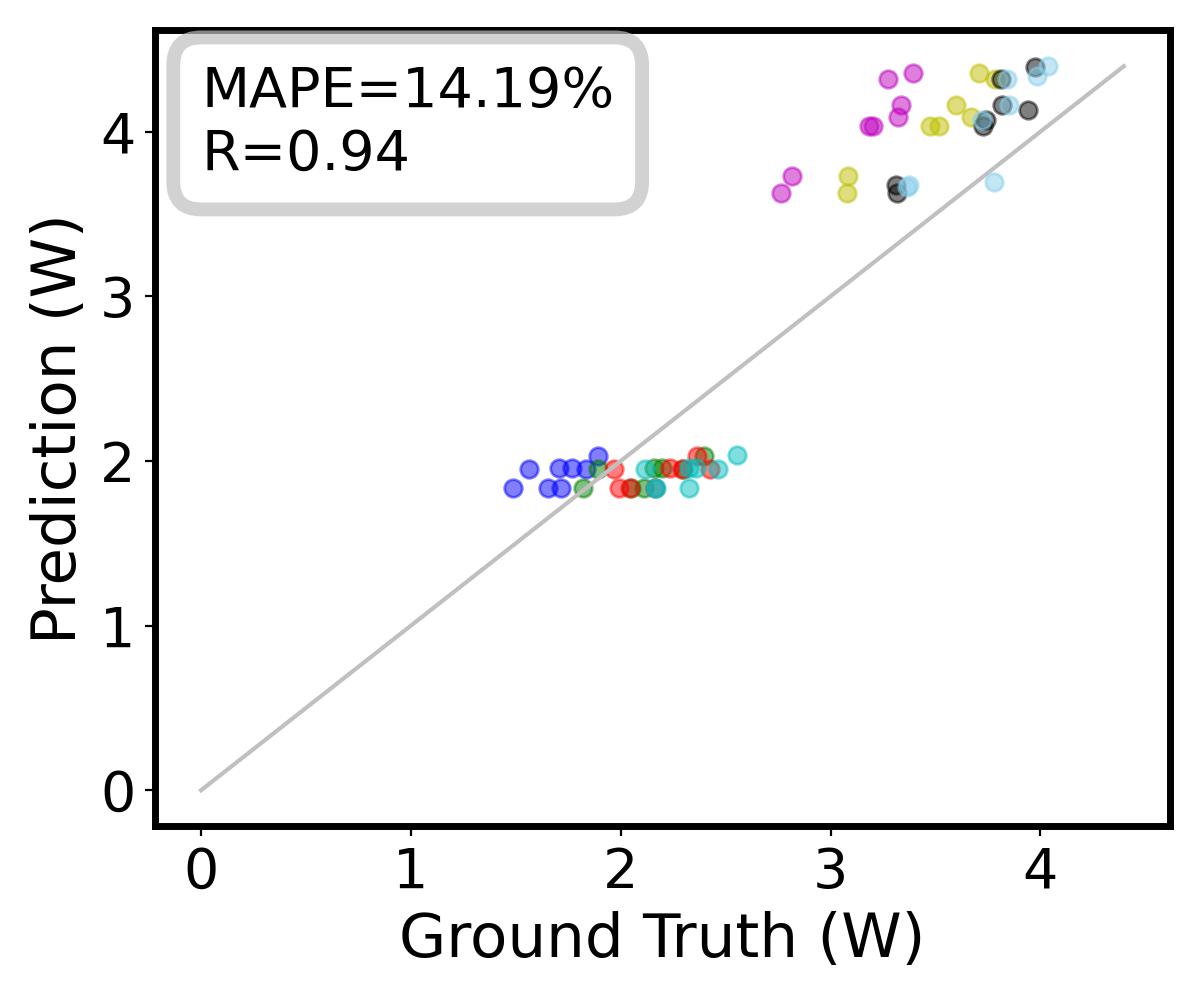}
    \label{mcpat_boom2xs2}
}

\vspace{-4mm}
%\hspace{-6mm}
\subfigure[FirePower (XS$\,\to\,$BOOM)]{
    \centering
    \includegraphics[width=0.20\textwidth]{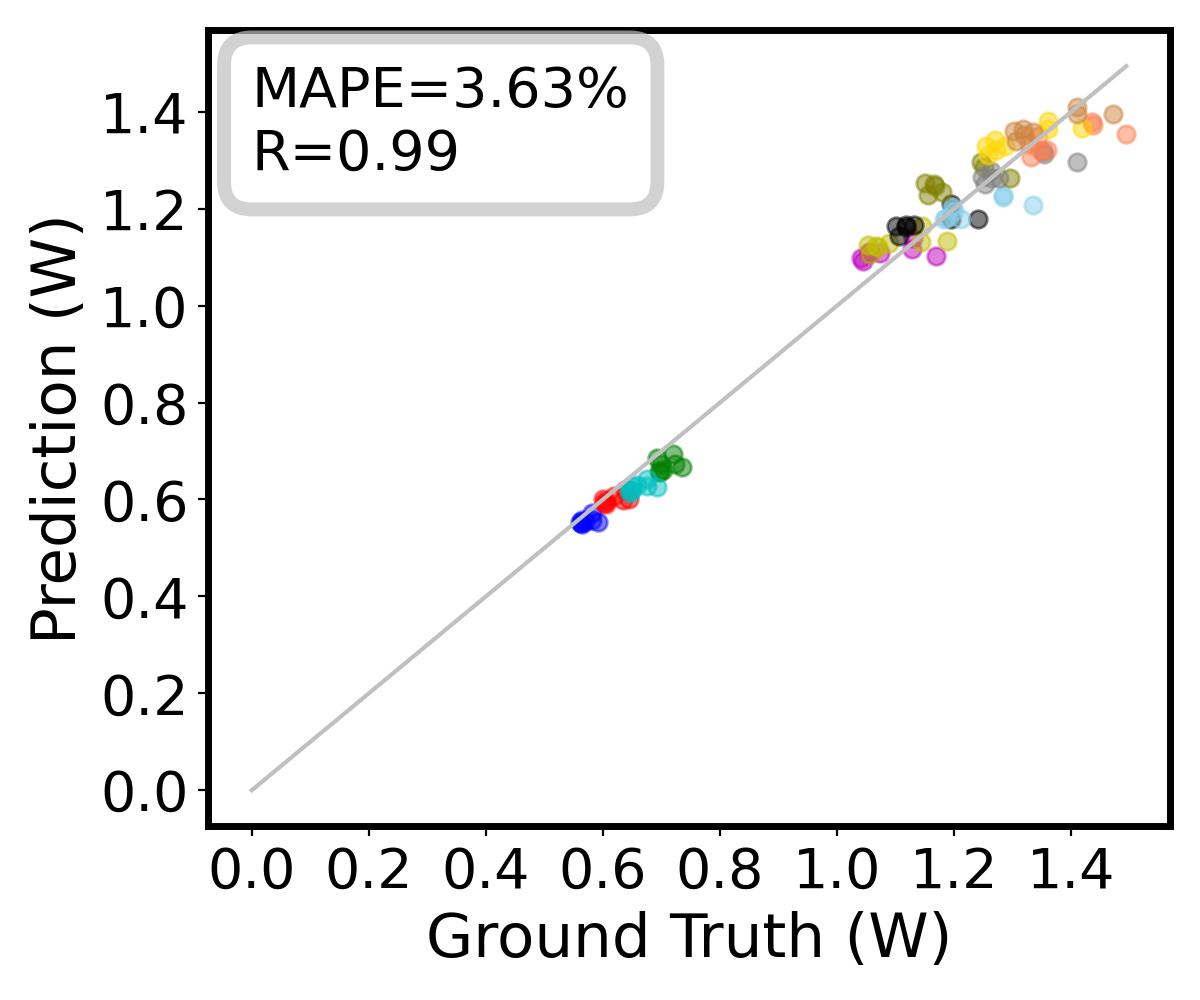}
    \label{FirePower_xs2boom2}
}
%\hspace{-2mm}
\subfigure[FirePower (BOOM$\,\to\,$XS)]{
    \centering
    \includegraphics[width=0.20\textwidth]{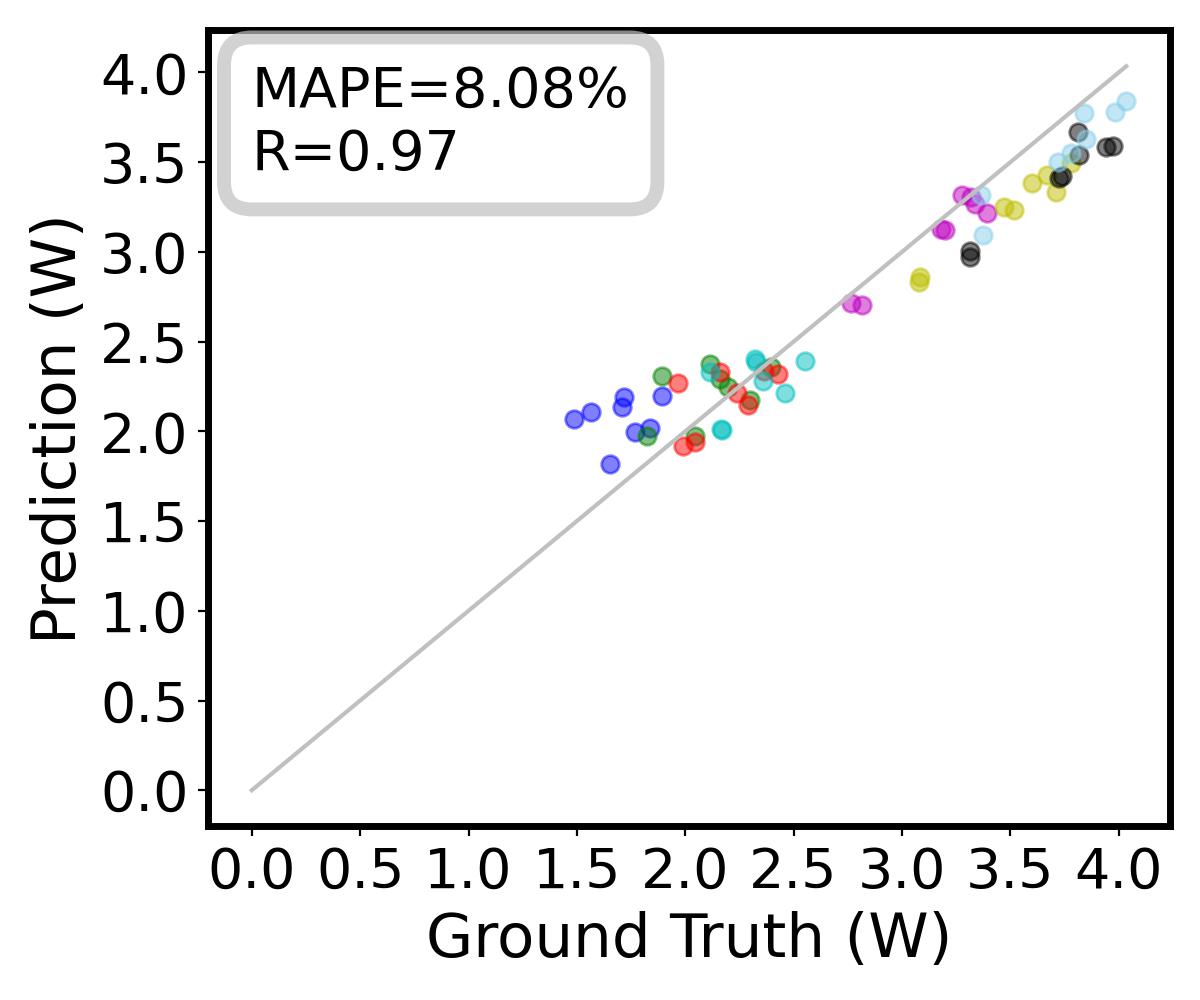}
    \label{FirePower_boom2xs2}
}
\vspace{-5mm}
\caption{Accuracy comparison between FirePower and McPAT-Calib (Available Config of Target Arch = 2).}
\vspace{-5mm}
\label{accuracy2}
\end{figure}

Fig.~\ref{ablation} summarizes comparisons between FirePower with our baseline, McPAT-Calib, and four ablation studies, under different numbers of available configurations of target architecture, i.e. 4, 3, and 2 configurations. Fig.~\ref{accuracy2} further visualize detailed results for FirePower and McPAT-Calib with only 2 available configurations, where samples of the same configuration are in the same color. The comparison with McPAT-Calib shows that FirePower can consistently achieve superior accuracy over McPAT-Calib regardless of scenarios. FirePower achieves the lowest MAPE and the highest correlation coefficient $R$, with at most (on average) 11.5\% (7\%) lower MAPE and 0.04 (0.03) higher correlation $R$ compared with McPAT-Calib. With only two configurations of target architecture, FirePower can still achieve a low MAPE of 5.8\% and a high correlation $R$ of 0.98 on average, which is 8.8\% lower in error percentage and 0.03 higher in $R$ compared with McPAT-Calib. The superiority of FirePower over McPAT-Calib is contributed by its ability to generalize knowledge acquired from known architecture. In contrast, architecture-specific McPAT-Calib trains model from scratch.

Fig.~\ref{ablation} also shows FirePower can constantly achieve the best accuracy compared with four ablation studies for both MAPE and correlation $R$.
McPAT-Calib + Component is an enhanced version of McPAT-Calib by building models for each component. It has an advantage over McPAT-Calib, validating the effect of power-friendly component definition. 
However, it can still not achieve a high accuracy compared with FirePower. It verifies knowledge generalization is critical to enable few-shot power modeling.

McPAT-Calib + Transfer Learning and McPAT-Calib + Component + Transfer Learning are two knowledge generalization methods based on transfer learning. They directly transfer the power model as a whole, regardless of generality. Results in Fig.~\ref{ablation} show that they outperform McPAT-Calib and McPAT-Calib + Component in many scenarios. It demonstrates that information from other known architecture can improve accuracy of few-shot modeling. 
But FirePower still has an obvious advantage over them, which is because these two methods do not decouple general and architecture-specific knowledge.
%where generalizing architecture-specific knowledge, i.e. knowledge related to event statistics, can have a negative impact on knowledge generalization. 
It verifies the necessity of decoupling generalizable and architecture-specific knowledge, where generalizing architecture-specific knowledge has negative impact. %the that only generalizing the generalizable knowledge is important.

Fig.~\ref{ablation} also shows that ``FirePower without Retraining'' is more accurate than McPAT-Calib + Component, also outperforming two transfer-learning-based methods in some scenarios.
This validates that even generalization with No Retraining can also work.
Comparison between FirePower and FirePower without Retraining verifies that the Retraining and strategy selection are crucial. The parameter importance is essential generalizable knowledge for FirePower.

%\vspace{-.05in}
\subsection{Generalization Evaluation}
\label{ger}
\begin{figure}[!t]
\centering
%\vspace{-6mm}
%\hspace{-4mm}
\subfigure[D-TLB (Accessible Config of Target)]{
    \centering
    \includegraphics[width=0.21\textwidth]{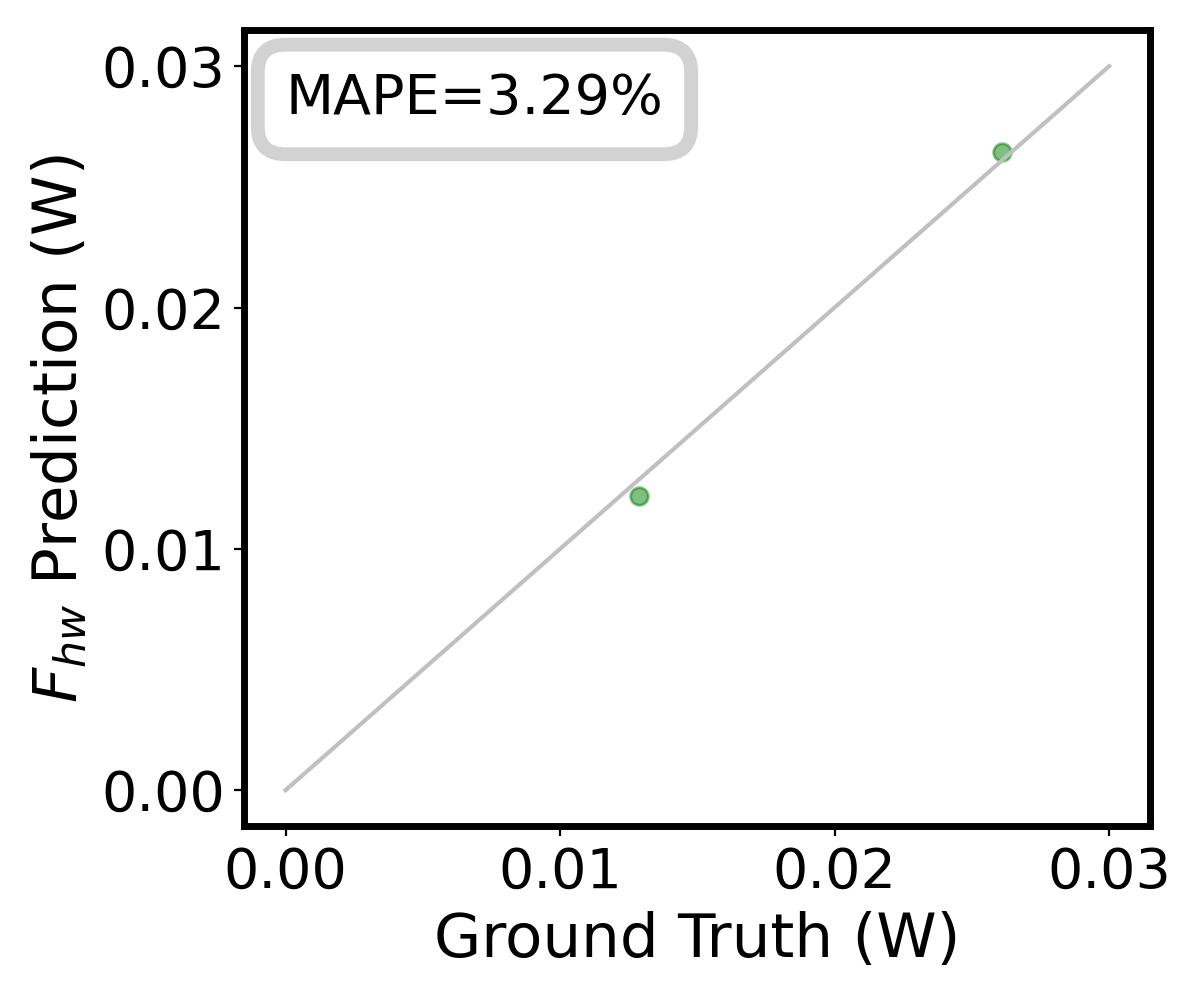}
}
\hspace{-1mm}
\subfigure[D-TLB (All Config of Target)]{
    \centering
    \includegraphics[width=0.21\textwidth]{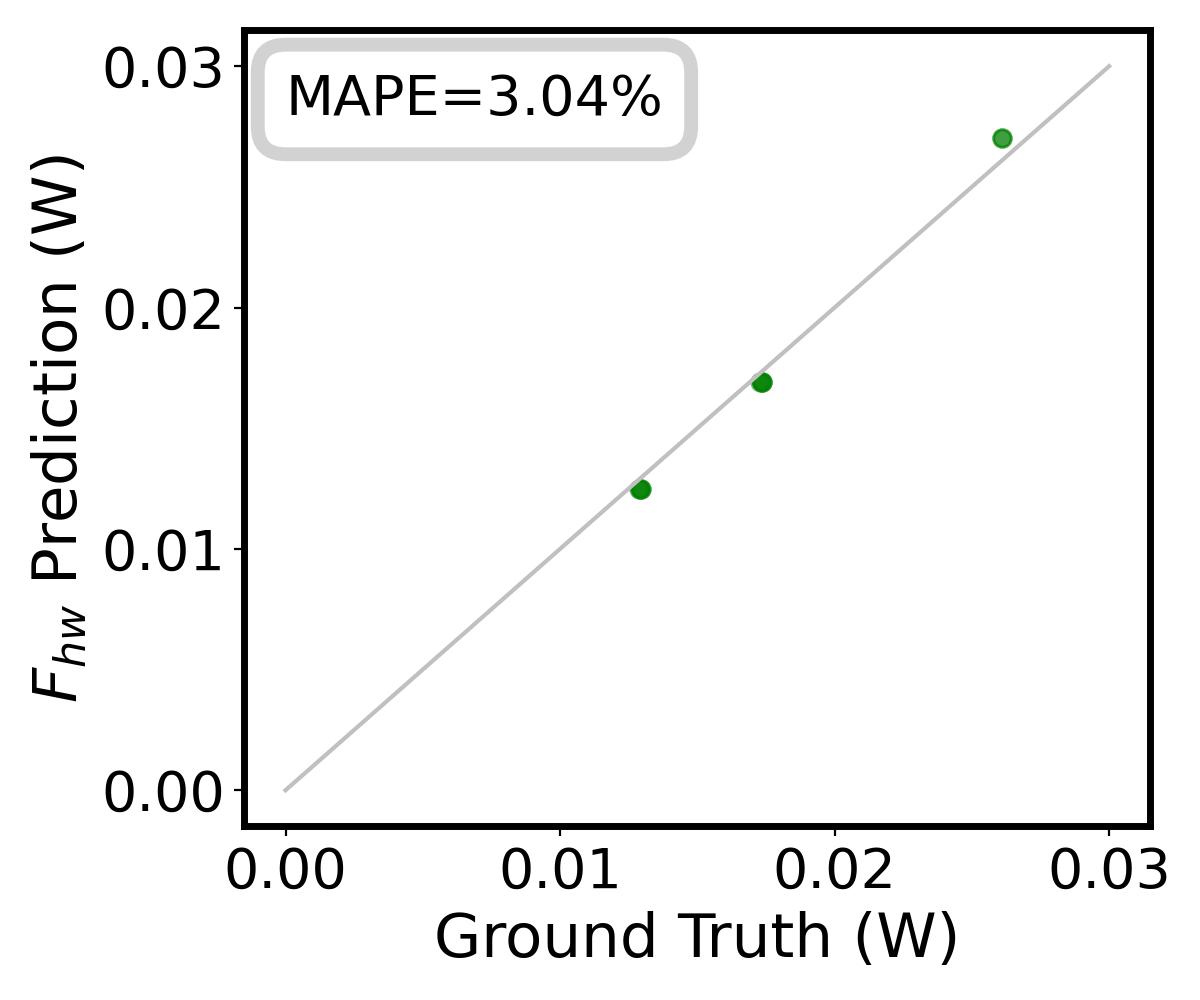}
}
\vspace{-3.5mm}

%\hspace{-4mm}
\subfigure[OtherLogic (Accessible Config of Target)]{
    \centering
    \includegraphics[width=0.21\textwidth]{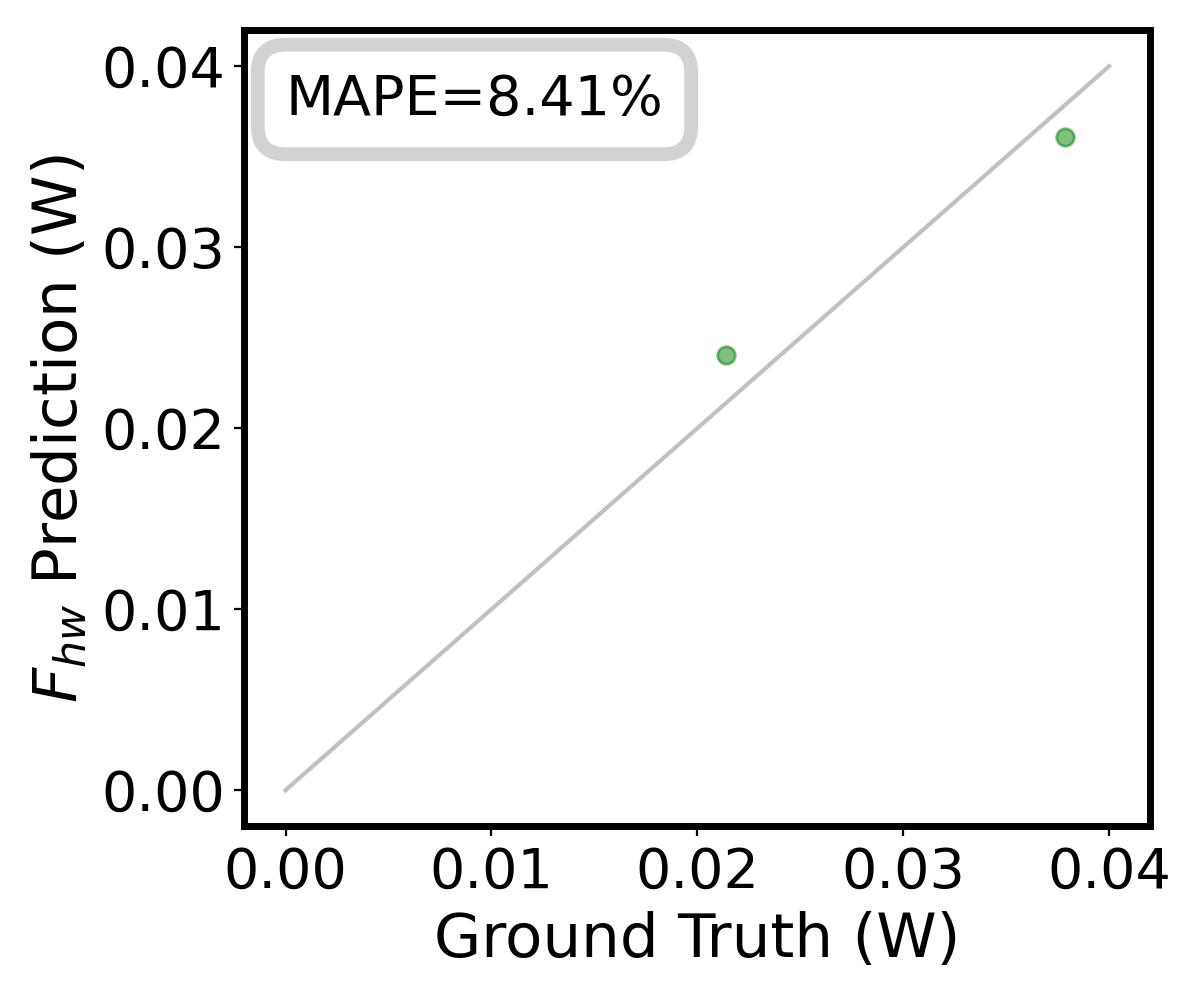}
}
\hspace{-1mm}
\subfigure[OtherLogic (All Config of Target)]{
    \centering
    \includegraphics[width=0.21\textwidth]{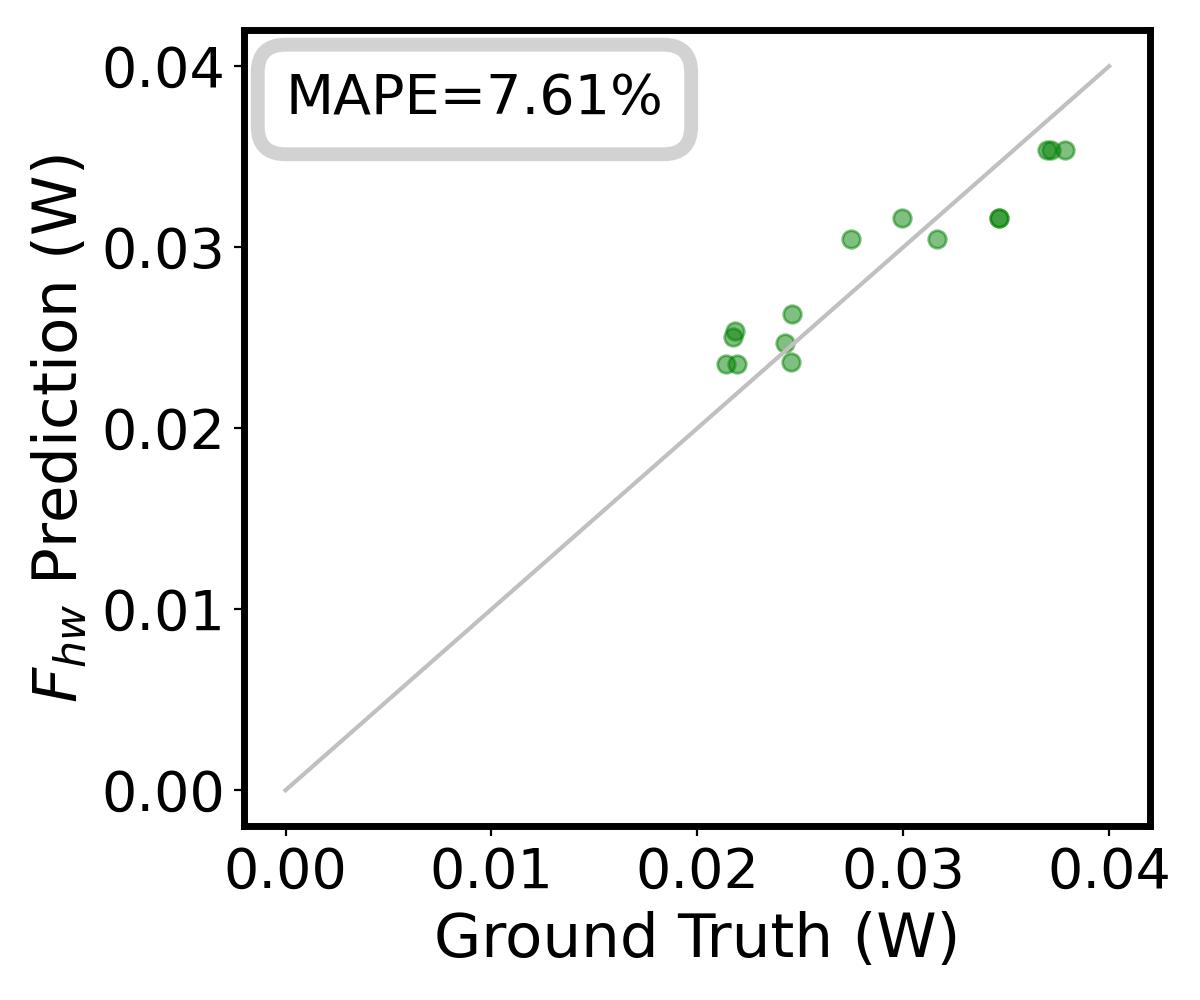}
}
\vspace{-6mm}
\caption{Generalization quality evaluation for components with high similarity across architectures. (a)(c) The generalization qualities observed with the accessible configuration of the target architecture. (b)(d) The generalization qualities evaluated with all configurations of the target architecture.}
\vspace{-4mm}
\label{highsimilarity}
\end{figure}

\begin{figure}[!t]
\centering
%\vspace{-2mm}

%\hspace{-4mm}
\subfigure[LSU (Accessible Config of Target)]{
    \centering
    \includegraphics[width=0.21\textwidth]{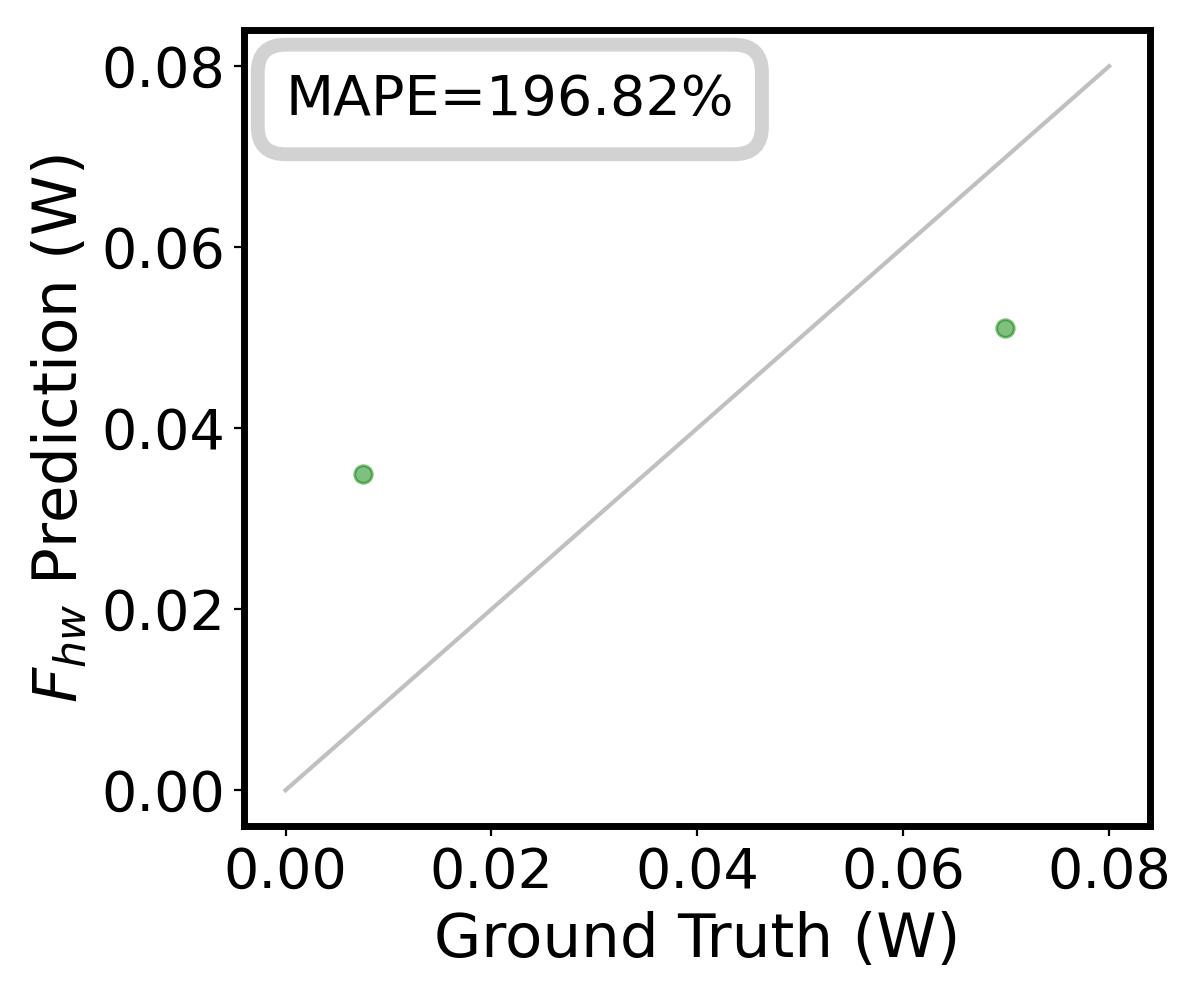}
}
\hspace{-1mm}
\subfigure[LSU (All Config of Target)]{
    \centering
    \includegraphics[width=0.21\textwidth]{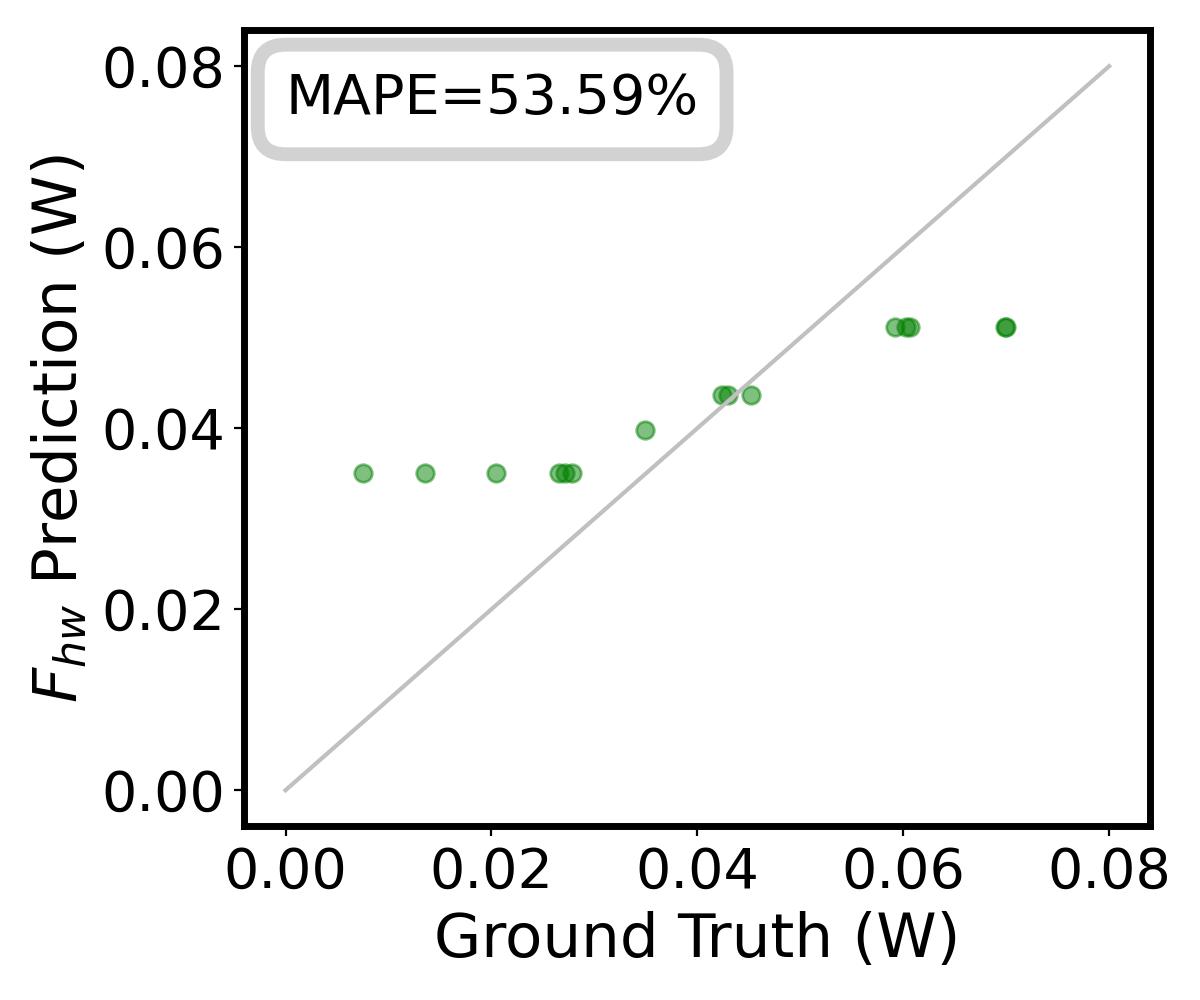}
}
\vspace{-5mm}
\caption{Generalization quality evaluation for components with low similarity across architectures. (a) The generalization quality observed with the accessible configuration of the target architecture. (b) The generalization quality evaluated with all configurations of the target architecture.}
\label{lowsimilarity}
\vspace{-6mm}
\end{figure}

%\vspace{-.03in}
Fig.~\ref{highsimilarity} illustrates the generalization evaluation for components with high similarity between known and target architecture. Fig.~\ref{lowsimilarity} shows the lower-similarity one. Higher similarity is supposed to result in a high generalization quality, and vice versa.

In Fig.~\ref{highsimilarity} and~\ref{lowsimilarity}, each point represents a configuration of the target architecture. The x-axis is the golden average power across workloads. The y-axis is the adjusted prediction of the hardware model trained on known architecture, which is discussed in Sec.~\ref{ge}. The similarity between these two values indicates generalization quality, which can be measured with MAPE.
Fig.~\ref{highsimilarity}(b)(d) and Fig.~\ref{lowsimilarity}(b) show actual generalization quality with all configurations of the target architecture, which is not accessible when building the model. Fig.~\ref{highsimilarity}(a)(c) and Fig.~\ref{lowsimilarity}(a) show generalization quality that we can observe with limited accessible configurations of target architecture.

We can find the actual generalization quality evaluated with all configurations in Fig.~\ref{highsimilarity}(b)(d) and Fig.~\ref{lowsimilarity}(b) correlates with generalization quality measured with the accessible ones in Fig.~\ref{highsimilarity}(a)(c) and Fig.~\ref{lowsimilarity}(a). This means architects can roughly estimate the quality with the generalization quality observed with the accessible configurations. Generally, as illustrated in these figures, if the generalization quality observed with the accessible configurations has a MAPE lower than 10\%, it suggests a relatively high actual generalization quality indicating that it can result in a high-quality generalized model. Conversely, if MAPE exceeds this threshold, the generalization may result in a low-quality model. In experiments, we always adopt generalized hardware model, because power percentage of evaluated low-similarity components is relatively small. 
%\input{_txt/5_discussion}
%\input{_txt/6_applications}

%\vspace{-.02in}
\section{Conclusion}
%\vspace{-.02in}
We propose FirePower which targets few-shot learning scenario for new target architectures by different users. Developer extracts the generalizable knowledge from a well-developed architecture, and then multiple projects can use this knowledge for few-shot power modeling, with limited available configurations of target architectures. The foundation-based paradigm reduces data requirement significantly, which is a compelling addition to architects’ toolbox.
%\vspace{-.02in}
\section*{Acknowledgement}
%\vspace{-.02in}
This work is partially supported by National Natural Science Foundation of China 62304192, and ACCESS – AI Chip Center for Emerging Smart Systems, sponsored by InnoHK funding, Hong Kong SAR. 
%We thank anonymous reviewers for their valuable feedback. 
We acknowledge the suggestions from Dr. Andrea Mondelli.
%\vspace{-.05in}

\bibliographystyle{ACM-Reference-Format}
\bibliography{refs}
%%%%%%%%%%%%%%%%%%%%%%%%%%%%%%%%%%%%

\end{document}